\documentclass{article} 
\usepackage{auxiliary/arxivpackages}
\usepackage{iclr2024_conference,times}

\iclrfinalcopy

\usepackage{auxiliary/mymacros}
\usepackage{auxiliary/commands}
\usepackage{auxiliary/filecommands}

\title{Efficient Inverse Multiagent Learning}

\author{Denizalp Goktas\thanks{Research conducted while the author was an intern at JP Morgan Chase \& Co.} \ \& Amy Greenwald  \\
Brown University, Computer Science\\
\texttt{denizalp\_goktas@brown.edu}
\AND
Sadie Zhao \\
Harvard University, Computer Science \\
\And
Alex Koppel \& Sumitra Ganesh \\
JP Morgan Chase \& Co\\
}

\begin{document}

\maketitle

\begin{abstract}
In this paper, we study inverse game theory (resp.\@ inverse multiagent learning) in which the goal is to find
parameters of a game's payoff functions for which the expected (resp.\@ sampled) behavior is an equilibrium. 
We formulate these problems as generative-adversarial (i.e., min-max) optimization problems, which we develop polynomial-time algorithms to solve, the former of which relies on an exact first-order oracle, and the latter, a stochastic one.
We extend our approach to solve inverse multiagent simulacral learning in polynomial time and number of samples.
In these problems, we seek a simulacrum, meaning parameters and an associated equilibrium that replicate the given observations in expectation.
We find that our approach outperforms the widely-used ARIMA method in predicting prices in Spanish electricity markets based on time-series data.

\end{abstract}

\section{Introduction}

Game theory provides a mathematical framework, called 
\mydef{games}, which is used to predict the outcome of the interaction of preference-maximizing agents called \mydef{players}. 
Each player in a game chooses a \mydef{strategy} from its \mydef{strategy space} according to its preference relation, often represented by a \mydef{payoff function} over possible \mydef{outcomes}, implied by a \mydef{strategy profile} (i.e., a collection of strategies, one-per-player). 
The canonical outcome, or \mydef{solution concept}, prescribed by game theory is the \mydef{Nash equilibrium (NE)} \citep{nash1950existence}: a strategy profile such that each player's strategy, fixing the equilibrium strategies of its opponents, is payoff-maximizing (or more generally, preference-maximizing).

In many applications of interest, such as contract design \citep{holmstrom1979moral, grossman1992analysis} and counterfactual prediction \citep{peysakhovich2019robust}, the payoff functions (or more generally, preference relations) of the players are not available, but the players' strategies are.
In such cases, we are concerned with estimating payoff functions 
for which these observed strategies are an equilibrium.
This estimation task serves to \mydef{rationalize} the players' strategies (i.e., we can interpret the observed strategies as solutions to preference-maximization problems).
Estimation problems of this nature characterize \mydef{inverse game theory} \citep{waugh2013computational, bestick2013inverse}.


The primary object of study of inverse game theory is the \mydef{inverse game}, which comprises a 
game with the 
payoff functions omitted, and an \mydef{observed strategy profile}.
The canonical solution concept prescribed for an inverse game is the \mydef{inverse Nash equilibrium}, i.e., payoff functions 
for which the observed strategy profile corresponds to a Nash equilibrium.
%
%
If the set of payoff functions in an inverse game is unrestricted, the set of inverse Nash equilibria can contain a wide variety of spurious solutions, e.g., in all inverse games, the payoff function that assigns zero payoffs to all outcomes is an inverse Nash equilibrium, because any strategy profile is a Nash equilibrium of a \mydef{constant game}: i.e., one whose payoffs are constant across strategies. 
To meaningfully restrict the class of payoff functions over which to search for an inverse Nash equilibrium, one common approach \citep{kuleshov2015inverse, syrgkanis2017inference} is to assume that the inverse game includes in addition to all the aforementioned objects, a \mydef{parameter-dependent payoff function} 
for each player, in which case an \mydef{inverse Nash equilibrium} is simply defined as parameter values such that the observed strategy profile is a Nash equilibrium of the parameter-dependent payoff functions evaluated at those values. 

If one assumes \emph{exact\/} oracle access to the payoffs of the game (i.e., if there exists a function which, for any strategy profile, returns the players' payoffs%
\footnote{Throughout this work, we assume that the oracle evaluations are constant time and measure computational complexity in terms of the number of oracle calls.}),
the problem of computing an inverse Nash equilibrium is one of \mydef{inverse multiagent planning}. 
In many games, however, a more appropriate assumption is \emph{stochastic\/} oracle access, because of inherent stochasticity in the game \citep{shapley1953stochastic} or because players employ randomized 
strategies \cite{nash1950existence}.
The problem of computing an inverse Nash equilibrium assuming stochastic oracle access is one of \mydef{inverse multiagent learning}.

One important class of inverse games is that of \mydef{inverse Markov games}, in which the underlying game is a \mydef{Markov game} \citep{shapley1953stochastic, fink1964equilibrium, takahashi1964equilibrium}, i.e., the game 
unfolds over an infinite time horizon: at each time period, players observe a state, take an action (simultaneously), receive a reward, and transition onto a new state. 
In such games, each player's strategy,%
\footnote{Throughout this paper, a strategy refers to the complete description of a players' behavior at any state or time of the game, while an action refers to a specific realization of a strategy at a given state and time.} 
also called a \mydef{policy}, is a mapping from states to actions describing the action the player takes at each state, with any strategy profile inducing a \mydef{history distribution} over \mydef{histories of play} i.e., sequences of (state, action profile) tuples. 
The payoff for any strategy profile is then given by its \mydef{expected cumulative reward} over histories of play drawn from the history distribution associated with the strategy profile. 
Excluding rare instances,%
\footnote{For simple enough games, one can express the expected cumulative reward in closed form, and then solve the inverse (stochastic) game  assuming exact oracle access.}
the payoff function in Markov games is only accessible via a stochastic oracle, typically implemented via a game simulator that returns \emph{estimates} of the value of the game's rewards and transition probabilities.
As such, the computation of an inverse Nash equilibrium in an inverse Markov game is an inverse multiagent learning problem, which is often called \mydef{inverse multiagent reinforcement learning (inverse MARL)} \citep{natarajan2010multi}.




In many real-world applications of inverse Markov games, such as robotics control \citep{coates2009apprenticeship}, one does not directly observe Nash equilibrium strategies but rather histories of play, which we assume are sampled from the history distribution associated with some Nash equilibrium. 
In these applications, we are given \mydef{an inverse simulation}---an inverse Markov game together with sample histories of play---based on which we seek parameter values which induce payoff functions that rationalize the observed histories. 
As a Nash equilibrium itself is not directly observed in this setting, we aim to compute parameter values that induce a Nash equilibrium that replicates the observed histories \emph{in expectation}.
We call the solution of such an inverse simulation (i.e., parameter values together with an associated Nash equilibrium) a \mydef{simulacrum}.
Not only does a simulacrum serve to explain (i.e., rationalize) observations, additionally, it can provide predictions of unobserved behavior.

We study two \mydef{simulacral learning} problems, a first-order version in which samples histories of play are faithful, and a second-order version in which they are not---a (possibly stochastic) function of each history of play is observed rather than the history itself.
Here, the use of the term ``first-order'' refers to the fact that the simulacrum does not necessarily imitate the actual equilibrium that generated the histories of play, 
since multiple equilibria can generate the same histories of play \citep{baudrillard1994simulacra}.
More generally, if the simulacrum is ``second-order,'' it is nonfaithful, meaning some information about the sample histories of play is lost.
We refer to the problems of computing first-order (resp.\@ second-order) simulacra as \mydef{first-order (resp.\@ second-order) simulacral learning}: i.e., build a first-order (resp.\@ second-order) simulacrum from faithful (resp.\@ non-faithful; e.g., aggregate agent behavior) sample histories of play.
We summarize the problems characterizing inverse game theory in \Cref{table:inverse_gt_summary}.

\vspace{-2.5mm}
\paragraph{Contributions}    


The algorithms introduced in this paper extend the class of games for which an inverse Nash equilibrium can be computed efficiently (i.e., in polynomial-time) to the class of normal-form concave games (which includes normal-form finite action games), finite state and action Markov games, and a large class of continuous state and action Markov games. 
While our focus is on Markov games in this paper, the results apply to normal-form \citep{nash1950existence}, Bayesian \citep{harsanyi1967games, harsanyi1968bayesian}, and extensive-form games \citep{zermelo1913anwendung}.
The results also extend to other equilibrium concepts, beyond Nash, such as (coarse) correlated \cite{aumann1974subjectivity, moulin1978strategically}, and more generally, $\Phi$-equilibrium \citep{greenwald2003general} \emph{mutatis mutandis}.





First, regarding inverse multiagent planning, we provide a min-max characterization of the set of inverse Nash equilibria of any inverse game for which the set of inverse Nash equilibria is non-empty, assuming an exact oracle (\Cref{thm:inverse_NE}).
We then show that for any inverse concave game, when the regret of each player is convex in the parameters of the inverse game, an assumption satisfied by a large class of inverse games such as inverse normal-form games, this min-max optimization problem is convex-concave, and can thus be solved in polynomial time (\Cref{thm:concave_game_inverse_NE}) via standard first-order methods.
This characterization also shows that the set of inverse Nash equilibria can be convex, even when the set of Nash equilibria is not.

Second, we generalize our min-max characterization to inverse multiagent learning, in particular inverse MARL, where we are given an inverse Markov game, and correspondingly, a \emph{stochastic\/} oracle, and we seek a first-order simulacrum (\Cref{thm:inverse_stoch_NE}).
We show that under standard assumptions, which are satisfied by a large class of inverse Markov games (e.g., all finite state and action Markov games and a class of continuous state and action Markov games), the ensuing min-max optimization problem is convex-gradient dominated, and thus an inverse Nash equilibrium can be computed once again via standard first-order methods in polynomial time (\Cref{thm:online_sgda}).

Third, we provide an extension of our min-max characterization to (second-order) simulacral learning (\Cref{thm:inverse_simulacrum}).
We once again characterize the problem as a solution to a min-max optimization problem, for which standard first-order methods compute a first-order stationary \citep{lin2020gradient} solution in polynomial-time, using a number of observations (i.e., unfaithful samples of histories of play) that is polynomial in the size of the inverse simulation (\Cref{thm:apprenticeship_thm}).

Finally, we include two sets of experiments.
In the first, we show that our method is effective in synthetic economic settings where the goal is to recover buyers' valuations from observed competitive equilibria (which, in this market, coincide with Nash equilibria).
Second, using real-world time-series data, we apply our method to predict prices in Spanish electricity markets, 
and find that it outperforms the widely-used ARIMA method in predicting prices on this real-world data set.

\begin{table}[t]
\begin{subtable}[t]{0.5\textwidth}
\resizebox{\columnwidth}{!}{%
\begin{tabular}{|l||c|c|}
\hline
Equilibrium Access & Exact Oracle & Stochastic Oracle \\
\hline \hline
Direct & Inverse Multiagent Planning & Inverse Multiagent Learning \\
\hline 
\makecell[l]{Faithful Samples} & First-order Simulacral Planning & First-order Simulacral Learning \\
\hline
\makecell[l]{Nonfaithful Samples} & Second-order Simulacral Planning & Second-order Simulacral Learning \\
\hline
\end{tabular}
}
\caption{Taxonomy of inverse 
game theory problems.
First-order simulacral learning is more commonly known as multiagent apprenticeship learning \citep{abbeel2004apprenticeship, yang2020inferring}.}
\label{table:inverse_gt_summary}
\vspace{-1em}
\end{subtable}
\begin{subtable}[t]{0.5\textwidth}
    \centering
    \resizebox{\columnwidth}{!}{%
    \begin{tabular}{|l|l|l|c|}
        \hline
        Reference & Game Type & Solution Concept & \makecell{Polytime?} \\
        \hline
        \hline \citep{fu2021evaluating} & Finite Markov & Nash & \xmark \\
        \hline \citep{yu2019multi} & Finite Markov & Quantal Response & \xmark \\
        \hline \citep{lin2019multi} & Finite Zero-sum Markov & Various & \xmark\\
        \hline 
        \citep{song2018multi} & Finite Markov &
        Quantal Response & \xmark \\
        \hline  \citep{syrgkanis2017inference} & Finite Bayesian & Bayes-Nash & \cmark \\
        \hline \citep{kuleshov2015inverse} & Finite Normal-Form & Correlated & \cmark \\
        \hline \citep{waugh2013computational} & Finite Normal-Form & Correlated &  \cmark \\
        \hline \citep{bestick2013inverse} & Finite Normal-Form & Correlated & \xmark \\  
        \hline \citep{natarajan2010multi} & Finite Markov & Cooperative & \xmark \\
        \hline \rowcolor{orange!50} This work & \makecell[l]{Finite/Concave Normal-form\\ Finite/Concave Markov} & 
        \makecell[l]{Nash/Correlated\\ Any Other \quad \quad \quad \quad }
        & \cmark \\
        \hline
    \end{tabular}
    }
    \caption{A comparison of our work and prior work on inverse game theory and inverse MARL.}
    \label{tab:summary_lit}
    \vspace{-2em}
\end{subtable}

\end{table}

\section{Preliminaries}
\vspace{-1em}
\textbf{Notation. } 
All notation for variable types, e.g., vectors, should be clear from context; if any confusion arises, see \Cref{sec_app:prelims}.
We denote 
by $[n]$ the set of integers $\left\{1, \hdots, n\right\}$.
Let $\calX$ be any set and $(\calX, \calF)$ any associated measurable space, where the $\sigma$-algebra $\calF$ unless otherwise noted is assumed to be the $\sigma$-algebra of countable sets, i.e., $\calF \doteq \{\calE \subseteq \calX \mid \calE \text{ is countable } \}$.
We write $\simplex(\calX) \doteq \{\mu: (\calX, \calF) \to [0, 1] \}$ to denote the set of  \mydef{probability measures} on $(\calX, \calF)$.
Additionally, we denote the orthogonal projection operator onto a set $\calX$ by $\project[\calX](\x) \doteq \argmin_{\y \in \calX} \left\|\x - \y \right\|_2^2$.
 

\textbf{Mathematical Concepts. } 
Consider any normed space $(\calX, \left\| \cdot \right\|)$ where $\calX \subset \R^\outerdim$ and any function $\obj: \calX \to \R$.
$\obj$ is $\lipschitz[\obj]$-\mydef{Lipschitz-continuous} w.r.t.\@ norm (typically, Euclidean)
$\left\| \cdot \right\|$ iff $\forall \x_1, \x_2 \in \calA, \left\| \obj(\x_1) - \obj(\x_2) \right\| \leq \lipschitz[\obj] \left\| \x_1 - \x_2 \right\|$.
If the gradient of $\obj$ is $\lipschitz[\grad \obj]$-Lipschitz-continuous, we refer to $\obj$ as $\lipschitz[\grad \obj]$-\mydef{Lipschitz-smooth}.
Furthermore, given $\scparam > 0$,  $\obj$ is said to be $\scparam$-\mydef{gradient-dominated} if $\min_{\x^\prime \in \calX} \obj(\x^\prime) \geq \obj(\x) + \scparam \cdot \min_{\x^\prime \in \calX} \left< \x^\prime - \x, \grad \obj(\x) \right>$ \citep{bhandari2019global}.

\amy{all throughout, in algos, appendix, etc., we have to check for $\actions$ vs.\@ $\actionspace$.}


\deni{Change One-shot to normal-form}

\amy{when/why was action changed to strategy? i noticed because it says ``an strategy'' all over the place}

\textbf{Normal-form Games. } 
A \mydef{(parametric)
game} $\game[][\param] \doteq (\numplayers, \numactions, \numparams, \stratspace, \params, \param, \util)$ 
comprises $\numplayers \in \N_+$ players, each $\player \in \players$ of whom chooses a strategy $\strat[\player] \in \stratspace[\player]$ from an strategy space $\stratspace[\player] \subseteq \R^{\numactions}$ simultaneously.
We refer to any vector of per-player strategies $\strat = (\strat[1], \hdots, \strat[\numplayers]) \in \stratspace$ as a \mydef{strategy profile}, where 
$\stratspace \doteq \bigtimes_{\player \in \players} \stratspace[\player] \subseteq \R^{\numplayers \numactions}$ denotes the space of all strategy profiles.
After the players choose their strategies $\strat \in \stratspace$, each receives a payoff $\util[\player](\strat; \param)$ given by payoff function $\util[\player]: \stratspace \times \params \to \R$ parameterized by a vector $\param$ in a parameter space $\params \subseteq \R^\numparams$.
We define the \mydef{payoff profile function} $\util(\strat; \param) \doteq \left( \util[\player](\strat; \param) \right)_{\player \in \players}$;
the \mydef{cumulative regret} $\cumulregret[][]: \stratspace \times \stratspace \times \params \to \R$ across all players, between two strategy profiles $\strat, \otherstrat \in \stratspace$, given $\param \in \params$, as $\cumulregret[][] (\strat, \otherstrat; \param) \doteq \sum_{\player \in \players} \util[\player](\otherstrat[\player], \strat[-\player]; \param) - \util[\player](\strat; \param)$;
and the \mydef{exploitability} (or \mydef{Nikaido-Isoda potential} \citep{nikaido1955note}) $\exploit (\strat; \param) \doteq \max_{\otherstrat \in \stratspace} \cumulregret[][] (\strat, \otherstrat; \param)$.

A game is said to be \mydef{concave} if for all parameters $\param \in \params$ and players $\player \in \players$, 1.~$\stratspace[\player]$ is non-empty, compact, and convex, 2.~$\util[\player]$ is continuous, and 3.~$\strat[\player] \mapsto \util[\player](\strat[\player], \strat[-\player]; \param)$ is concave.
Given $\param \in \params$, an $\varepsilon$-\mydef{Nash equilibrium} ($\varepsilon$-NE) of a game $\game[][\param]$ is a strategy profile 
$\strat[][][][*] \in \stratspace$ s.t.\ $\util[\player](\strat[][][][*]; \param) \geq \max_{\strat[\player] \in \stratspace[\player]} \util[\player](\strat[\player], \strat[-\player][][][*]; \param) - \varepsilon$, for all players $\player \in \players$.
A $0$-Nash equilibrium is simply called a Nash equilibrium, and
is guaranteed to exist in concave games \citep{nash1950existence, arrow-debreu}. 

\textbf{Dynamic Games. } An \mydef{(infinite-horizon, discounted, parametric) Markov game} \citep{shapley1953stochastic, fink1964equilibrium, takahashi1964equilibrium} $\mgame[][\param] \doteq (\numplayers, \numactions, \states, \actionspace, \params, \param, \reward, \trans, \discount, \initstates)$ is a dynamic game played over an infinite time horizon.
The game initiates at time $\iter = 0$ in some state $\staterv[0] \sim \initstates$ 
drawn from an \mydef{initial state distribution} $\initstates \in \simplex (\states)$.
At each time period $\iter = 0, 1, \hdots$, each player $\player \in \players$ plays an \mydef{action} $\action[\player][][\iter] \in \actionspace[\player]$ from an action space $\actionspace[\player] \subset \R^\numactions$.
We define the space of action profiles $\actionspace = \bigtimes_{\player \in \players} \actionspace[\player]$.
After the players choose their \mydef{action profile} $\action[][][\iter] \doteq (\action[1][][\iter], \hdots, \action[\numplayers][][\iter]) \in \actionspace$, each player $\player$ receives a \mydef{reward}
$\reward[\player] (\state[\iter][], \action[][][\iter]; \param)$ according to a \emph{parameterized} \deni{Removed footnote}
\mydef{reward profile function} $\rewards: \states \times \actionspace \times \params \to \R^\numplayers$.
The game then either ends with probability $1-\discount$, where $\discount \in (0,1)$ is called the \mydef{discount factor}, or transitions to a new state $\staterv[\iter+1] \sim \trans(\cdot \mid \state[\iter], \action[][][\iter])$ according to a \mydef{(Markov) probability transition kernel} $\trans$ whereby for all $(\state, \action) \in \states \times \actionspace$, $\trans (\cdot \mid \state, \action) \in \simplex(\states)$, and $\trans (\state[\iter+1] \mid \state[\iter], \action[][][\iter]) = \Pr(\staterv[\iter + 1] = \state[\iter+1] \mid \staterv[\iter] = \state[\iter], \actionrv[][][\iter] = \action[][][\iter])$ is the probability of transitioning to state $\state[\iter+1]$ from state $\state[\iter]$ when the players' take action profile $\action[][][\iter]$.%
\footnote{For notational convenience, 
we assume the probability transition function is independent of the parameters, but we note that our min-max characterizations apply more broadly without any additional assumptions, while our polynomial-time computation results apply when, in addition to \Cref{assum:convex_param_stoch}, one assumes the probability transition function is stochastically convex (see, for instance, \citet{atakan2003stochastic}) in the parameters of the game.}

A \mydef{(stationary Markov) policy} \citep{maskin2001markov} for player $\player \in \players$ is a mapping $\policy[\player]: \states \to \actionspace$ from states to actions so that $\policy[\player](\state) \in \actionspace[\player]$ denotes the action that player $\player$ takes at state $\state$.
For each player $\player \in \players$, we define the space of all (measurable) policies $\policies[\player] \doteq \left\{\policy[\player]: \states \to \actionspace[\player]\right\}$. 
As usual, $\policy \doteq (\policy[1], \hdots, \policy[\numplayers]) \in \policies \doteq \smash{\bigtimes_{\player \in \players} \policies[\player]}$ denotes a \mydef{policy profile}.
A \mydef{history (of play)} $\hist[][][] \in (\states \times \actionspace)^\numiters$ of length $\numiters \in \N$ is a sequence of state-action tuples $\hist[][][] = (\state[\iter], \action[][][\iter])_{\iter = 0}^{\numiters - 1}$.  
For any policy profile $\policy \in \policies$,
define the \mydef{(discounted) history distribution} $\histdistrib[][\policy] (\hist[][][]) \doteq \initstates (\state[0]) \prod_{\iter = 0}^\numiters \discount^\iter \trans (\state[\iter +1] \mid \state[\iter], \policy (\state[\iter]))$ as the probability of observing a history $\hist$ of length $\numiters$. 
Throughout, we denote by $\histrv[][] \doteq \left( \staterv[\iter], \actionrv[][][\iter] \right)_\iter \sim \histdistrib[][\policy]$ any randomly sampled history from $\histdistrib[][\policy]$.\footnote{Let $(\states, \calF_{\states})$, $(\actionspace, \calF_{\actionspace})$, and $(\states \times \actionspace, \calF_{\states \times \actionspace})$ be the measurable spaces associated with the state, action profile, and state-action profile ($\states \times \actionspace$) spaces, respectively. 
Further, let $([0,1], \calB_{[0,1]})$, $(\R^n, \calB_{\R^n})$ be measurable spaces on $[0,1]$ and $\R^n$ defined by the Borel $\sigma$-algebra. 
For simplicity, we do not explicitly represent the reward profile function, transition probability kernel, initial state distribution, or policies as measures or measurable functions.
We note, however, that for the expectations we define to be well-posed, 
they all must be assumed to be measurable functions.
We simply write $\reward: \states \times \actionspace \to \R^\numplayers$, $\trans: \states \times (\states \times \actionspace) \to [0,1]$, $\initstates: \states \to [0,1]$, and $\policy: \states \to \actionspace$ to mean, respectively, $\reward: (\states \times \actionspace, \calF_{\states \times \actionspace}) \to (\R^\numplayers, \calB_{\R^\numplayers})$, $\trans: (\states, \calF_{\states}) \times (\states \times \actionspace, \calF_{\states \times \actionspace}) \to ([0,1], \calB_{[0,1]})$, $\initstates: (\states, \calF_{\states}) \to ([0,1], \calB_{[0,1]})$, and $\policy: (\states, \calF_{\states}) \to (\actionspace, \calF_{\actionspace})$.}

Fix a policy profile $\policy \in \policies$ and a player $\player$.
In our analysis of Markov games, we rely on the following terminology. 
The \mydef{expected cumulative payoff} 
is given by $\util[\player](\policy; \param) \doteq \Ex_{\histrv \sim \histdistrib[][\policy]} \left[\sum_{\iter = 0}^\infty \reward[\player](\staterv[\iter], \actionrv[][][\iter]; \param) \right].$
The \mydef{state-} 
and \mydef{action-value functions}  \
are defined, respectively, as
$\vfunc[\player][\policy] (\state; \param) \doteq \Ex_{\histrv \sim \histdistrib[][\policy]} \left[\sum_{\iter = 0}^\infty  \reward[\player] (\staterv[\iter], \actionrv[][][\iter]; \param) \mid \staterv[0] = \state \right]$
and
$\qfunc[\player][\policy] (\state, \action; \param) \doteq \Ex_{\histrv \sim \histdistrib[][\policy]} \left[\sum_{\iter = 0}^\infty  \reward[\player](\staterv[\iter], \actionrv[][][\iter]; \param) \mid \staterv[0] = \state, \actionrv[][][0] = \action \right]$. 
The \mydef{state occupancy distribution}
$\statedist[\initstates][\policy] \in \simplex(\states)$ denotes the probability that a state is reached under a policy $\policy$, given initial state distribution $\initstates$, i.e., $\statedist[\initstates][\policy] (\state) \doteq  \Ex_{\histrv \sim \histdistrib[][\policy]} \left[ \sum_{\iter = 0}^\infty \setindic[{\staterv[\iter] = \state}] \right]$. 
%
Finally, as usual, an $\varepsilon$-\mydef{Nash equilibrium} ($\varepsilon$-\nash) of a game $\mgame[][\param]$ is a policy profile $\policy[][][*] \in \policies$ such that for all $\player \in \players$, $\util[\player] (\policy[][][*]; \param) \geq \max_{\policy[\player] \in \policies[\player]} \util[\player] (\policy[\player], \policy[-\player][][*]; \param) - \varepsilon$; and a Nash equilibrium ensues when $\varepsilon = 0$.\deni{Removed fn here.}%

\section{Inverse Multiagent Planning}
\label{sec:inverse_planning}

The goal of inverse multiagent planning is to invert an equilibrium: i.e., estimate a game's parameters, given observed behavior. 
In this section, we present our main idea, namely a zero-sum game (i.e., min-max optimization) characterization of inverse multiagent planning, where one player called the stabilizer picks parameters, while the other called the destabilizer picks per-player deviations.
This game is zero-sum) because the stabilizer seeks parameters that rationalize (i.e., minimize the exploitability of) the observed equilibrium, while the destabilizer aims to rebut the rationality of the observed equilibrium (i.e., seeks deviations that maximize cumulative regret).
We use this characterization to develop a gradient descent ascent algorithm that finds inverse NE in polynomial time, assuming access to \mydef{an exact first-order oracle}: specifically, a pair of functions that return the value and gradient of the payoff profile function.

An \mydef{inverse game} $\game[][-1] \doteq 
(\game[][\paramtrue] \setminus \paramtrue, \truestrat)$ comprises a \mydef{game form} (i.e., a parametric game \emph{sans\/} its parameter) $\game[][\paramtrue] \setminus \paramtrue$ together with an observed strategy profile $\truestrat$, which we assume is a Nash equilibrium.
Crucially, we do not observe the parameters $\paramtrue$ of the payoff functions.
%
%
Given an inverse game $\game[][-1]$, our goal is to compute an \mydef{$\varepsilon$-inverse Nash equilibrium}, meaning parameter values $\param[][][*] \in \params$ s.t.\@ $\truestrat \in \stratspace$ is an $\varepsilon$-NE of $\game[][{\param[][][*]}]$.
As usual, a 0-inverse NE is simply called an inverse NE.
Note that this definition does not require that we identify
the true parameters $\paramtrue$, as identifying $\paramtrue$ is impossible unless there exists a bijection between the set of parameters and the set of Nash equilibria, a highly restrictive assumption that is not even satisfied in games with a unique Nash equilibrium.
To compute an inverse NE is to find parameter values that minimize the exploitability of the observed equilibrium.
This problem is a min-max optimization problem, as the parameter values that minimize exploitability are those that maximize the players' cumulative regrets.
More precisely:

\begin{restatable}{theorem}{thminverseNE}
\label{thm:inverse_NE}
    The set of inverse NE of $\game[][-1]$ is the set of parameter profiles $\param \in \params$ that solve the optimization problem $\min_{\substack{\param \in \params}} \exploit (\truestrat; \param)$, or equivalently, this min-max optimization problem:
    \begin{align} 
        \min_{\substack{\param \in \params}} \max_{\otherstrat \in \stratspace} \obj (\param, \otherstrat) \doteq 
        \cumulregret[] (\truestrat, \otherstrat; \param) = \sum_{\player \in \players} \left[\util[\player] (\otherstrat[\player], \truestrat[-\player]; \param) - \util[\player] (\truestrat; \param) \right]
        \label{eq:min_max_gen_sim}
    \end{align}
\vspace{-1.5em}
\end{restatable}

\amy{Generalization of inverse optimization to game-theoretic settings? do you cite \href{https://homes.cs.washington.edu/~todorov/papers/DvijothamICML10.pdf}{this paper}, or others by Todorov?}

This min-max optimization problem can be seen as a generalization of the dual of \citeauthor{waugh2013computational}'s (\citeyear{waugh2013computational}) maximum entropy likelihood maximization method for games with possibly continuous strategy spaces, taking Nash equilibrium rather than maximum entropy correlated equilibrium as the inverse equilibrium. 
In contrast to \citeauthor{waugh2013computational}'s dual, our min-max optimization problem characterizes the set of \emph{all\/} inverse NE, and not only a subset of the inverse correlated equilibria, in particular those that maximize entropy.
This formulation also generalizes \citeauthor{swamy2021moments}'s (\citeyear{swamy2021moments}) moment matching game from a single-agent to a multiagent setting.

%
\begin{wrapfigure}{L}{0.533\textwidth}
\vspace{-1cm}
\begin{minipage}{0.533\textwidth}
\begin{algorithm}[H]
\caption{Adversarial Inverse Multiagent Planning}
\textbf{Inputs:} $\params, \stratspace, \obj, \learnrate[\param],  \learnrate[\otherstrat], \numiters, \param[][0], \otherstrat[][][0], \truestrat$  \\
\textbf{Outputs:} $(\param[][\iter], \otherstrat[][][\iter])_{t = 0}^\numiters$
\label{alg:gda}
\begin{algorithmic}[1]
\For{$\iter = 0, \hdots, \numiters - 1$}
    
    \State  $\param[][\iter + 1] \gets \project[\params] \left[\param[][\iter] - \learnrate[\param][\iter] \grad[{\param}] \obj(\param[][\iter], \otherstrat[][][\iter]) \right]$
    
    \State  $\otherstrat[][][\iter + 1]  \gets \project[\stratspace] \left[ \otherstrat[][][\iter] + \learnrate[\otherstrat][\iter] \grad[{\otherstrat}] \obj(\param[][\iter], \otherstrat[][][\iter]) \right]$

\EndFor
\State \Return $(\param[][\iter], \otherstrat[][][\iter])_{t = 0}^\numiters$
\end{algorithmic}
\end{algorithm}
\end{minipage}
\vspace{-0.33cm}
\end{wrapfigure}

Without further assumptions, the objective function $\obj$ in \Cref{eq:min_max_gen_sim} is non-convex non-concave; however, 
under suitable assumptions (\Cref{assum:concave_game}) satisfied by
finite action normal-form games, for example, it becomes convex-concave.

\amy{how natural is the convex regret assumption? how often does it follow assuming concave utilities?} \deni{It works in finite-action games games.} \amy{let's make sure this is explained somewhere.}

\begin{assumption}
\label{assum:concave_game}
    Given an inverse game $\game[][-1]$, assume 1.~(Concave game) for all parameters $\param \in \params$, $\game[][\param]$ is concave; and 
    2.~(Convex parametrization) $\params$ is non-empty, compact, and convex; and for all $\forall \player \in \players$, $\otherstrat[\player] \in \stratspace[\player]$, and $\truestrat \in \stratspace$, each player $\player$'s regret $\param \mapsto \util[\player] (\otherstrat[\player], \truestrat[-\player]; \param) - \util[\player] (\truestrat; \param)$ is convex.
\end{assumption}



\begin{remark}
\label{rem:convex}
    Perhaps surprisingly, the set of inverse NE can be convex even when the set of NE is not, since the set of solutions to a convex-concave (or even convex-non-concave)\deni{removed footnote} 
    min-max optimization problem is convex. 
    \sdeni{More concretely, while in normal-form games the set of Nash equilibria is non-convex beyond highly structured classes such as zero-sum, potential, and monotone games, the set of inverse Nash equilibria is convex for all normal-form games.}{}
    This observation should alleviate any worries about the computational intractability of inverse game theory that might have been suggested by the computational intractability of game theory itself \citep{daskalakis2009complexity, chen2006settling}. 
\end{remark}

If additionally, we assume the players' payoffs are Lipschitz-smooth (\Cref{assum:smoothness}), \Cref{eq:min_max_gen_sim} can then be solved to $\varepsilon$ precision in $O\left( \nicefrac{1}{\varepsilon^2} \right)$ via gradient descent ascent (\Cref{alg:gda}). 
That is, as \Cref{thm:concave_game_inverse_NE} shows, an inverse $\varepsilon$-NE can be computed in $O\left( \nicefrac{1}{\varepsilon^2} \right)$ iterations.%
\footnote{We include detailed theorem statements and proofs in \Cref{sec:app_proofs}.}
We note that this convergence complexity can be further reduced to $O\left( \nicefrac{1}{\varepsilon} \right)$ (even without decreasing step-sizes) if one instead applies an extragradient descent ascent method \citep{golowich2020eglast} or optimistic GDA \citep{gorbunov2022last}.

\begin{assumption}[Lipschitz-Smooth Game]
\label{assum:smoothness}
    For all players $\player \in \players$, $\util[\player]$ is $\lipschitz[{\grad \util[\player]}]$-Lipschitz-smooth.
\end{assumption}


\begin{restatable}[Inverse NE Complexity]{theorem}{thmconcavegameinverseNE}
\label{thm:concave_game_inverse_NE}
    Under Assumptions \ref{assum:concave_game}--\ref{assum:smoothness}, for $\varepsilon \geq 0$, if \Cref{alg:gda} is run with inputs that satisfy $\numiters \in \Omega(\nicefrac{1}{\varepsilon})$ and for all $\iter \in [\numiters]$, $\learnrate[\otherstrat][\iter] = \learnrate[\param][\iter] \asymp \nicefrac{1}{\lipschitz[\grad \obj]}$, 
    then the time-average of all parameters $\mean[{\param[][\numiters]}] \doteq \frac{1}{\numiters + 1}\sum_{\iter = 0}^\numiters \param[][\iter]$ is an $\varepsilon$-inverse NE.
\end{restatable}

\section{Inverse Multiagent Reinforcement Learning}
\label{sec:inverse_marl}

In this section, we build on our zero-sum game (i.e., min-max optimization) characterization of inverse game theory to tackle inverse MARL in an analogous fashion.
As it is unreasonable to assume exact oracle access to the players' (cumulative) payoffs in inverse MARL, we relax this assumption
in favor of a stochastic oracle model.
More specifically, we assume access to a \mydef{differentiable game simulator} \citep{suh2022differentiable}, which simulates histories of play $\hist \sim \histdistrib[][\policy]$ according to $\histdistrib[][\policy]$, given any policy profile $\policy$, and returns the rewards $\reward$ and transition probabilities $\trans$,%
\footnote{We note that in inverse reinforcement learning, as opposed to reinforcement learning, it is typical to assume that the transition model is known (see, for instance \citep{abbeel2004apprenticeship}, Footnote 8).} 
encountered along the way, together with their gradients.
\sdeni{We can then average across multiple simulations to estimate estimate deviation payoffs, cumulative regrets, and the requisite gradients, to recover the parameters $\param[][][*]$ for which $\truepolicy$ is a Nash equilibrium 
using a policy gradient algorithm that converges 
in polynomial time.}{}

Formally, an \mydef{inverse Markov game} $\mgame[][-1] \doteq
(\mgame[][\paramtrue] \setminus \paramtrue, \truepolicy)$ is an inverse game that comprises a \mydef{Markov game form} (i.e., a parametric Markov game \emph{sans\/} its parameter) $\mgame[][\paramtrue] \setminus \paramtrue$ together with an observed policy profile $\truepolicy$, which we assume is a Nash equilibrium.
Crucially, we do not observe the parameters $\paramtrue$ of the payoff functions.
Since a Markov game is a normal-form game with payoffs given by $\util(\policy; \param) = \Ex_{\histrv \sim \histdistrib[][\policy]} \left[\sum_{\iter = 0}^\infty \reward(\staterv[\iter], \actionrv[][][\iter]; \param) \right]$, the usual definitions of inverse NE and cumulative regret apply, and the following result, which characterizes the set of inverse NE as the minimizers of a \emph{stochastic\/} min-max optimization problem, is a corollary of \Cref{thm:inverse_NE}.

\begin{corollary}
    \label{thm:inverse_stoch_NE}
    The set of inverse NE of $\mgame[][-1]$ is characterized by solutions to the following
    problem: 
    \begin{align}     
        \min_{\substack{\param \in \params}} \max_{\policy \in \policies} \obj (\param, \policy) \doteq
        \sum_{\player \in \players}  \Ex_{\substack{\histrv \sim \histdistrib[][{(\policy[\player], \truepolicy[-\player])}]\\ \histrv[][\dagger] \sim \histdistrib[][\truepolicy]}} \left[\sum_{\iter = 0}^\infty \reward[\player](\staterv[\iter], \actionrv[][][\iter]; \param) - 
        \sum_{\iter = 0}^\infty \reward[\player](\staterv[\iter][][\dagger], \actionrv[][][\iter][\dagger]; \param) \right] 
    \label{eq:inverse_stoch_NE}
    \end{align}
\end{corollary}

\amy{i think that perhaps all the $x$'s in this section should be changed to $y$'s, to match the notation in the previous section. but this is a dangerous change to attempt? so maybe what we should do is just change all the $y$'s to $x$'s in the SGDA algo?}\deni{I agree but I think let's do this for the Arxiv version.}

As is usual in \sdeni{deep}{}reinforcement learning, we use policy gradient to solve the destabilizer's problem in \Cref{eq:inverse_stoch_NE}.
To do so, we restrict the destabilizer's action space to a policy class $\policies[][\stratspace]$ parameterized by $\stratspace \subset \R^\numstrats$.
Redefining $\obj (\param, \strat) \doteq \obj (\param, \policy[][][\strat])$, for $\policy[][\strat] \in \policies[][\stratspace]$, we aim to solve the stochastic min-max optimization problem $\min_{\param \in \params} \max_{\strat \in \stratspace} \obj (\param, \strat)$. 
Solutions to this problem are a superset \amy{subset? actually, i'm not sure why they intersect at all?} \deni{superset, because if the policy class becomes less representative, then the destabilizer will become ``weaker'', and the stabilizer will ``win'' more often. They intersect because restricting the policy class only makes the action space of the destabilizer smaller, but anything that was in the larger action space is still part of the solutions. This is kinda like the intuition of how CE is CCE because the set of swap deviations is larger than the set external deviations.} of the solutions to \Cref{eq:inverse_stoch_NE},
unless it so happens that all best responses can be represented by policies in $\policies[][\stratspace]$, \deni{Deleted footnote}
because restricting the expressivity of the policy class decreases the power of the destabilizer.
As in \Cref{sec:inverse_planning}, without any additional assumptions, $\obj$ is in general non-convex, non-concave, and non-smooth.
While we can ensure convexity and smoothness of $\param \mapsto \obj (\param, \strat)$ under suitable assumptions on the game parameterization, namely by assuming the regret at each state is convex in $\param$, concavity in $\strat$ is not satisfied even by finite state and action Markov games. 
Under the following conditions, however, we can guarantee that $\obj$ is Lipschitz-smooth,
convex in $\param$, and gradient dominated in $\strat$.


\begin{assumption}[Lipschitz-Smooth Gradient-Dominated Game]
\label{assum:smooth_convex_invex}
    Given an inverse Markov game $\mgame[][-1]$, assume
    1.~$\states$ and $\actionspace$ are non-empty, and compact;
    2.~(Convex parameter spaces) $\stratspace, \params$ are non-empty, compact, and convex; 
    3.~(Smooth Game) $\grad \reward$, $\grad \trans$, and $\grad[\strat] \policy[][][\strat]$, \amy{do we already assume earlier (or know) that $\reward, \trans, \policy[][][\strat]$ are differentiable?} \deni{No, I don't think so?} \amy{i guess i just don't understand why the gradients have to be differentiable if the functions themselves don't have to be? why can't we use sub-gradients of the gradients as well?} 
    for all policies $\policy[][][\strat] \in \policies[][\stratspace]$, are continuously differentiable;
    4.~(Gradient-Dominated Game) for all players $\player \in \players$, states $\state \in \states$, action profiles $\action \in \actionspace$, and policies $\policy[][][\strat] \in \policies[][\stratspace]$, $\strat \mapsto \qfunc[\player][{\policy[][][\strat]}](\state, \policy[][][\strat](\state); \param)$ is $\scparam$-gradient-dominated for some $\scparam>0$; and
    5.~(Closure under Policy Improvement)
    for all states $\state \in \states$, players $\player \in \players$, and policy profiles $\policy \in \policies$, there exists $\policy[][][\outer] \in \policies[][\outerset]$ s.t.\@ $\qfunc[\player][\policy] (\state, \policy[\player][][\outer] (\state), \policy[-\player] (\state)) = \max_{\policy[\player][][\prime] \in \policies[\player]} \qfunc[\player][\policy] (\state, \policy[\player][][\prime] (\state), \policy[-\player] (\state))$.
\end{assumption}


%
Part 3 of \Cref{assum:smooth_convex_invex} implies that the game's cumulative payoff function is Lipschitz-smooth in the policy parameters $\strat$.
We note that a large class of Markov games satisfy Part 4, including 
linear quadratic games \citep{bhandari2019global}, 
finite state and action games, and continuous state and action games whose rewards (resp.\@ transition probabilities) are concave (resp.\@ stochastically concave) in each player's action \citep{atakan2003valfunc}. 
Finally, Part 5 is a standard assumption (see, for instance, Section 5 of \citet{bhandari2019global}), which guarantees that the policy parameterization is expressive enough to represent best responses.

\begin{assumption}[Convex Parameterization]
\label{assum:convex_param_stoch}
    Given an inverse Markov game $\mgame[][-1]$, assume that for all players $\player \in \players$, states $\state \in \states$, and action profiles $\action, \otheraction \in \actionspace$, the per-state regret $\param \mapsto \reward[\player](\state, \otheraction[\player], \action[-\player]; \param) - \reward[\player](\state, \action; \param)$ is convex.
\end{assumption}

With these assumptions in hand, we face a convex gradient-dominated optimization problem, i.e., $\param \mapsto \obj (\param, \strat)$ is convex, for all $\strat \in \stratspace$, and $\strat \mapsto \obj (\param, \strat)$ gradient-dominated, for all $\param \in \params$. 
As for normal-form games (see \Cref{rem:convex}), the set of inverse NE in Markov games is convex under Assumptions~\ref{assum:smooth_convex_invex} and \ref{assum:convex_param_stoch}.
Consequently, we can obtain polynomial-time convergence of stochastic gradient descent ascent (\Cref{alg:online-sgda}) by slightly modifying known results \citep{daskalakis2020independent}.


\begin{wrapfigure}{L}{0.6125\textwidth}
    \vspace*{-0.8cm}
    \begin{minipage}{0.6125\textwidth}
    \begin{algorithm}[H]
    \caption{Adversarial Inverse MARL} 
    \textbf{Inputs:} $\params, \policies, \obj[\param], \obj[\strat], \learnrate[\param][ ], \learnrate[\strat][ ], \numiters, \param[][0], \strat[][][0], \truepolicy$ \\
    \textbf{Outputs:} $(\param[][\iter], \strat[][][\iter])_{t = 0}^\numiters$
    \label{alg:online-sgda}
    \begin{algorithmic}[1]
    \For{$\iter = 0, \hdots, \numiters - 1$}
            
            \State 
            $\histmatrix \sim \bigtimes_{\player \in \players} \histdistrib[][{(\policy[\player][][{\strat[][][\iter]}], \truepolicy[-\player])}]$, $\hist[][\dagger][] \sim \histdistrib[][\truepolicy]$
            
            \State $\param[][\iter + 1] \gets \project[\params] \left[\param[][\iter] - \learnrate[\param][\iter] \obj[\param] (\param[][\iter], \strat[][][\iter]; \histmatrix, \hist[][\dagger]) \right]$
        
            \State $\strat[][][\iter + 1] \gets \project[\policies] \left[ \strat[][][\iter] + \learnrate[\strat][\iter] \obj[\strat] (\param[][\iter], \strat[][][\iter]; \histmatrix, \hist[][\dagger]) \right]$
    
            
    \EndFor
    \State \Return $(\param[][\iter], \strat[][][\iter])_{t = 0}^\numiters$
    \end{algorithmic}
    \end{algorithm}
    \end{minipage}
    \vspace*{-0.4cm}
\end{wrapfigure}


\Cref{alg:online-sgda} requires an estimate of $\grad \obj$ w.r.t.\@ both $\param$ and $\strat$.
Under Part 3 of \Cref{assum:smooth_convex_invex}, the gradient of $\obj$ w.r.t.\@ $\strat$ can be obtained by the deterministic policy gradient theorem \citep{silver2014deterministic}, while the gradient of $\obj$ w.r.t.\@ $\param$ can be obtained by the linearity of the gradient and expectation operators.
However, both of these gradients involve an expectation---over $\histrv \sim \histdistrib[][{(\policy[\player][][\strat], \truepolicy[-\player])}]$ and $\histrv[][\dagger] \sim \histdistrib[][\truepolicy]$.
As such, we estimate them using simulated trajectories from the deviation 
history distribution $\histmatrix \doteq \left(\hist[][1], \hdots, \hist[][\numplayers]\right)^T \sim \bigtimes_{\player \in \players} \histdistrib[][{(\policy[\player][][\strat], \truepolicy[-\player])}]$ \amy{can we delete the $T$ superscript? isn't it implied?} and 
the equilibrium history distribution $\hist[][\dagger][] \sim \histdistrib[][\truepolicy]$, respectively.
For a given such pair $(\histmatrix, \hist[][\dagger])$, the cumulative regret gradient estimators $\obj[\param]$ and $\obj[\strat]$ correspond to the gradients 
of the cumulative regrets between each deviation history $\hist[][\player]$ in $\histmatrix$ and $\hist[][\dagger]$, and can be computed directly using the chain rule for derivatives, as we assume access to a differentiable game simulator.%
\footnote{For completeness, we show how to compute $\obj[\strat]$ and $\obj[\param]$ in \Cref{sec_app:gradient_estimate}.
In our experiments, however, as has become common practice in the literature \citep{mora2021pods}, we compute these gradients by simply autodifferentiating the cumulative regret of any history w.r.t.\@ the policy parameters using a library like Jax \citep{jax2018github}. 
We also show that under \Cref{assum:smooth_convex_invex}, $(\obj[\param], \obj[\strat])$ is an unbiased estimate of $(\grad[\param]  \obj, \grad[\strat] \obj)$ whose variance is bounded. \amy{is this notation incorrect? feels like you should have something like $\hat{\nabla}$ before each $f$.} \deni{We recently switched to that type of notation but I think let's keep it this way for the camera-ready and we can change for the Arxiv submission?} \amy{well then what about $(\hat{\obj[\param]}, \hat{\obj[\strat]})$?} \deni{We don't have macros for them. Seems dangerous.}}


Finally, we define the \mydef{equilibrium distribution mismatch coefficient} $\|\nicefrac{\partial\statedist[\initstates][{\policy[][][\dagger]}]}{\partial \initstates} \|_\infty$ as the Radon-Nikodym derivative of the state occupancy distribution of the NE $\policy[][][\dagger]$ w.r.t.\@ the initial state distribution $\initstates$.
This coefficient, which measures the inherent difficulty of reaching states under $\policy[][][\dagger]$, 
is closely related to other distribution mismatch coefficients introduced in the analysis of policy gradient methods \citep{agarwal2020optimality}. 
With this definition in hand, we can finally show polynomial-time convergence of stochastic GDA (\Cref{alg:online-sgda}) under Assumptions \ref{assum:smooth_convex_invex}--\ref{assum:convex_param_stoch}.

\begin{theorem}
\label{thm:online_sgda}
    Under Assumptions \ref{assum:smooth_convex_invex}--\ref{assum:convex_param_stoch}, for all $\varepsilon \in (0,1)$, if \Cref{alg:online-sgda} is run with inputs that satisfy $\numiters \in \Omega\left( \varepsilon^{-10}\|\nicefrac{\partial\statedist[\initstates][{\policy[][][\dagger]}]}{\partial \initstates} \|_\infty\right)$ and for all $\iter \in [\numiters]$, $\learnrate[\otherstrat][\iter] \asymp  \varepsilon^4$ and
    $\learnrate[\param][\iter] \asymp \varepsilon^8$,   
    then the time-average of all parameters $\mean[{\param[][\numiters]}] \doteq \frac{1}{\numiters + 1}\sum_{\iter = 0}^\numiters \param[][\iter]$ is an $\varepsilon$-inverse NE.
\end{theorem}

\section{Simulacral Learning}






\deni{The content in the next two paragraphs feels there but organized slightly weirdly, maybe?} \amy{what do we need to say beyond: don't observe eqm, just trajectories. after inferring parameter values, we can push them forward to generate eqm policies. in this sense, an ensuing eqm is part of the solution -- my main problem with this, btw, is: what if the ensuing eqm is not unique? i guess you don't care. you'll take any eqm that generates the observed behavior.}

\amy{it seems to me that we are brushing under the rug a key assumption, namely that of the ``observation distribution''. although we cannot simulate policies per se, this ``observation distribution'' effectively acts like a simulator, given observational info about policies. so i don't think the result in this section is as magical as we seem to be making it out to be.}

In this section, we consider the more realistic setting in which we do not observe an equilibrium, but observe only sample histories $\left\{ \hist[][][\numsample] \right\}_{\numsample} = \left\{(\state[\iter, \numsample], \action[][][\iter, \numsample])_\iter \right\}_{\numsample} \sim \histdistrib[][\truepolicy]$ associated with an \emph{unobserved\/} equilibrium $\truepolicy$. 
The problem of interest then becomes one of not only inferring parameter values from observed behavior, but of additionally finding equilibrium policies that generate the observed behavior, a solution which we refer to as a first-order simulacrum.
A first-order simulacrum can be seen as a generalization of an inverse equilibrium, as it not only comprises parameters that rationalize the observed histories, but also policies that mimic them in expectation.
First-order simulacral learning is also known as \mydef{multiagent apprenticeship learning} \citep{abbeel2004apprenticeship, yang2020inferring}.

\amy{a simulacrum is generative. so i don't think it is a disaster to call the whole setup, and even the paper, generative-adversarial, even though in the earlier sections we do not use the generative capabilities of this, our most general, problem and solution.} \deni{I don't think a simulacrum is generative in the traditional sense of the world, i.e., it is not a stochastic function?}

\begin{wrapfigure}{l}{0.66\textwidth}
    \vspace{-0.8cm}
    \begin{minipage}{0.66\textwidth}
    \begin{algorithm}[H]
    \caption{Adversarial Simulacral Learning}
    \textbf{Inputs:} $\params, \policies, (\avg[{\otherobj[\param]}], \avg[{\otherobj[\strat]}], \avg[{\otherobj[\otherstrat]}]), \learnrate[\param][ ], \learnrate[\strat][ ], \learnrate[\otherstrat][ ], \numiters, \param[][0], \strat[][][0], \otherstrat[][][0], \{\trueobs[][\numsample]\}$ \\
    \textbf{Outputs:} $(\param[][\iter], \strat[][][\iter], \otherstrat[][][\iter])_{t = 0}^\numiters$
    \label{alg:offline-sgda}
    \begin{algorithmic}[1]
    \For{$\iter = 0, \hdots, \numiters - 1$}
            
            \State 
            $\histmatrix \sim \bigtimes_{\player \in \players} \histdistrib[][{(\policy[\player][][{\strat[][][\iter]}], \policy[-\player][][{\otherstrat[][][\iter]}])}]$, $\hist[][][] \sim \histdistrib[][{(\policy[][][{\strat[][][\iter]}])}]$
            
            \State $\param[][\iter + 1] \gets \project[\params] \left[\param[][\iter] - \learnrate[\param][\iter] \grad[\param]  \avg[{\otherobj[\param]}] (\param[][\iter], \strat[][][\iter], \otherstrat[][][\iter]; \histmatrix, \hist[][]) \right]$

            \State $\strat[][][\iter+1] \gets \project[\policies] \left[ \strat[][][\iter] - \learnrate[\strat][\iter] \grad[\strat] \avg[{\otherobj[\strat]}] (\param[][\iter], \strat[][][\iter],  \otherstrat[][][\iter]; \histmatrix, \hist[][]) \right]$
            
            \State $\otherstrat[][][\iter+1] \gets \project[\policies] \left[ \otherstrat[][][\iter] + \learnrate[\otherstrat][\iter] \grad[\otherstrat] \avg[{\otherobj[\otherstrat]}] (\param[][\iter], \strat[][][\iter],  \otherstrat[][][\iter]; \histmatrix, \hist[][]) \right]$
    
            
    \EndFor
    \State \Return $(\param[][\iter], \strat[][][\iter], \otherstrat[][][\iter])_{t = 0}^\numiters$
    \end{algorithmic}
    \end{algorithm}
    \end{minipage}
    \vspace{-0.5cm}
\end{wrapfigure}

Even more generally, we might not have access to samples $\{ \hist[][\dagger][\numsample]\}_{\numsample \in [\numsamples]} \sim \histdistrib[][\truepolicy]$ from an equilibrium history distribution, but rather a lossy function of those histories according to some function $\rho: \hists \to \obspace$ that produces \mydef{observations} $\{ \trueobs[][\numsample] \}_{\numsample \in [\numsamples]} \doteq \{ \bm{\rho} \left(\hist[][\dagger][\numsample]\right)\}_{\numsample \in [\numsamples]} \sim \obsdistrib[][][\truepolicy]$, distributed according to some \mydef{(pushforward) observation distribution} 
$\obsdistrib[][][{\policy}] \in \simplex(\obspace)$, parameterized by policy profile $\policy \in \policies$, where $\obspace$ is the observation space.
This more general framework is very useful in applications where there are limitations on the data collection process: e.g., if there are game states at which some of the players' actions are unobservable, or when only an unfaithful function of them is available.
Here, we seek to learn the more general notion of a \mydef{second-order simulacrum}.



\deni{Explain difference between online and offline, bc for a given parameters, player's best response is not optimal.}

Formally, an \mydef{inverse simulation} $\inversesim \doteq (\mgame[][\paramtrue] \setminus \paramtrue, \obspace, \obsdistrib, \obsdistrib[][][\truepolicy])$ is a tuple consisting of a Markov game form $\mgame[][\paramtrue] \setminus \paramtrue$ with unknown parameters $\paramtrue$, an observation 
distribution $\obsdistrib: \policies \to \simplex (\obspace)$ mapping policies to distributions over the \mydef{observation space} $\obspace$, and an observation distribution $\obsdistrib[][][\truepolicy]$ for the \emph{unobserved\/} behavioral policy $\truepolicy$, which we assume is a Nash equilibrium.
%
%
%
Our goal is to find an \mydef{$(\varepsilon, \delta)$-Nash simulacrum}, meaning a tuple of parameters and policies $(\param[][][*], \policy[][][*]) \in \params \times \policies$ that $(\varepsilon, \delta)$-\mydef{simulates} the observations as a Nash equilibrium: i.e., 
$\util[\player] (\policy[][][*]; \param[][][*]) \geq \max_{\policy[\player] \in \policies[\player]} \util[\player] (\policy[\player], \policy[-\player][][*]; \param[][][*]) - \varepsilon $ and $\Ex_{(\obs,\trueobs) \sim \obsdistrib[][][{\policy[][][*]}] \times \obsdistrib[][][\truepolicy]} \left[\left\|\obs - \trueobs[]\right\|^2 \right] \leq \delta$.
%
%
\Cref{thm:inverse_simulacrum}, which is analogous to \Cref{thm:inverse_stoch_NE}, characterizes the set of Nash simulacra of an inverse simulation. 

\begin{restatable}{theorem}{thminversesimulacrum}
\label{thm:inverse_simulacrum}
Given an inverse simulation $\inversesim$,
    for any $\mixparamone, \mixparamtwo > 0$, the set of Nash simulacra of $\mgame[][-1]$ is equal to the set of minimizers of the following stochastic min-max optimization problem:
    \begin{align}
        \min_{\substack{\param \in \params \\ \policy \in \policies}} \exploit(\param, \policy) = \min_{\substack{\param \in \params \\ \policy \in \policies}} \max_{\otherpolicy \in \policies} \otherobj(\param, \policy, \otherpolicy) \doteq  \mixparamone \Ex_{\substack{(\obs,\trueobs) \sim \obsdistrib[][][\policy] \times \obsdistrib[][][\truepolicy]}} \left[\left\|\obs - \trueobs \right\|^2 \right] + \mixparamtwo \cumulregret(\policy, \otherpolicy; \param)\label{eq:min_max_simulacrum}
    \end{align}
\end{restatable}

To tackle simulacral learning, we approximate $\otherobj$ via realized observation samples $\{ \trueobs[][\numsample] \} \sim \obsdistrib[][][\truepolicy]$, based on which we compute the empirical learning loss $\avg[\otherobj](\param, \policy, \otherpolicy) \doteq \mixparamone \Ex_{\substack{\obs \sim \obsdistrib[][][\policy]}} \left[\nicefrac{1}{\numsamples} \sum_{\numsample = 1}^\numsamples \left\|\obs - \trueobs[][\numsample] \right\|^2 \right]$ $+$ $\mixparamtwo \cumulregret(\policy, \otherpolicy; \param)$.
Additionally, as in the previous section, we once again restrict policies to lie within a parametric class of policies $\policies[][\stratspace]$, redefine $\otherobj(\param, \strat, \otherstrat) \doteq \otherobj(\param, \policy[][][\strat], \policy[][][\otherstrat])$ and $\avg[\otherobj](\param, \strat, \otherstrat) \doteq \avg[\otherobj](\param, \policy[][][\strat], \policy[][][\otherstrat])$, and solve the ensuing optimization problem over the empirical learning loss $\min_{(\param, \strat) \in \params \times \stratspace} \max_{\otherstrat \in \stratspace} \avg[\otherobj](\param, \strat, \otherstrat)$.


In general, this stochastic min-max optimization is non-convex non-concave. 
By \Cref{assum:smooth_convex_invex}, however, the function $\otherstrat \mapsto \otherobj (\param, \strat, \otherstrat)$ is gradient dominated, for all $\param \in \params$ and $\strat \in \stratspace$.
Nevertheless, it is not possible to guarantee that $(\param, \strat) \mapsto \otherobj (\param, \strat, \otherstrat)$ is convex or gradient dominated, for all $\otherstrat \in \otherstratspace$, without overly restrictive assumptions.
This claim is intuitive, since the computation of an inverse simulacrum involves computing a Nash equilibrium policy, which in general is a PPAD-complete problem \citep{daskalakis2009complexity,foster2023hardness}.  
Finally, defining gradient estimators as we did in \Cref{sec:inverse_marl}, to obtain gradient estimators $(\avg[{\otherobj[\param]}], \avg[{\otherobj[\strat]}], \avg[{\otherobj[\otherstrat]}])(\param, \strat, \otherstrat; \histmatrix, \hist[][\strat])$ from samples histories $\histmatrix \sim \bigtimes_{\player \in \players} \histdistrib[][{(\policy[\player][][\strat], \policy[-\player][][\otherstrat])}]$ and $\hist[][\strat][] \sim \histdistrib[][{\policy[][][\strat]}]$,\amy{drop $\strat$ superscript on $\hist$? (twice)} we can use \Cref{alg:offline-sgda} to compute a local solution of \Cref{eq:min_max_simulacrum} from polynomially-many observations.

\begin{restatable}{theorem}{thmapprenticeshipthm}\label{thm:apprenticeship_thm}
    Suppose that \Cref{assum:smooth_convex_invex} holds, and that for all $\policy[][][\strat] \in \policies[][\stratspace]$ , $\obsdistrib[][][{\policy[][][\strat]}]$ is twice continuously differentiable in $\strat$.
    For any $\varepsilon \in (0,1)$, if \Cref{alg:offline-sgda} is run with inputs that satisfy $\numiters \in \Omega\left(\nicefrac{\variance^2}{ \varepsilon^{10}}\|\nicefrac{\partial\statedist[\initstates][{\policy[][][*]}]}{\partial \initstates} \|_\infty\right)$ and for all $\iter \in [\numiters]$, $\learnrate[\otherstrat][\iter] \asymp  \varepsilon^4$\samy{}{, $\learnrate[\strat][\iter] \asymp  \varepsilon^8$,} and $\learnrate[\param][\iter] \asymp \varepsilon^8$,
    then the best iterate  $\smash{(\bestiter[{\param}], \bestiter[{\strat}][])}$ 
    converges to an $\varepsilon$-stationary point 
    of $\exploit$ (defined in \Cref{sec:app_proofs}).
     Additionally, for any $\zeta, \xi \geq 0$, it holds with probability $1-\zeta$ that
     $
            \avg[{\exploit}](\bestiter[{\param}][\numiters], \bestiter[{\strat}][\numiters]) - \exploit(\bestiter[{\param}][\numiters], \bestiter[{\strat}][\numiters]) \leq \xi
    $ if the number of sample observations $\numsamples \in \Omega(\nicefrac{1}{\xi^2} \log(\nicefrac{1}{\zeta}))$.
    \vspace{-1em}
\end{restatable}

\section{Experiments}
\vspace{-1em}
We run two sets of experiments with the aim of answering two questions. 
Our first goal is to understand the extent to which our algorithms are able to compute inverse Nash equilibria, if any, beyond our theoretical guarantees.
Our second goal is to understand the ability of game-theoretic models to make predictions about the future.%
\footnote{Our code can be found \href{https://anonymous.4open.science/r/Generative-Adversarial-Inverse-Multiagent-Learning-ICLR2024-1C2C/}{here}. \amy{EXPIRED!}}

\begin{wrapfigure}{R}{0.4\textwidth}
    \centering
    \vspace{-1cm}
    \includegraphics[scale=0.5]{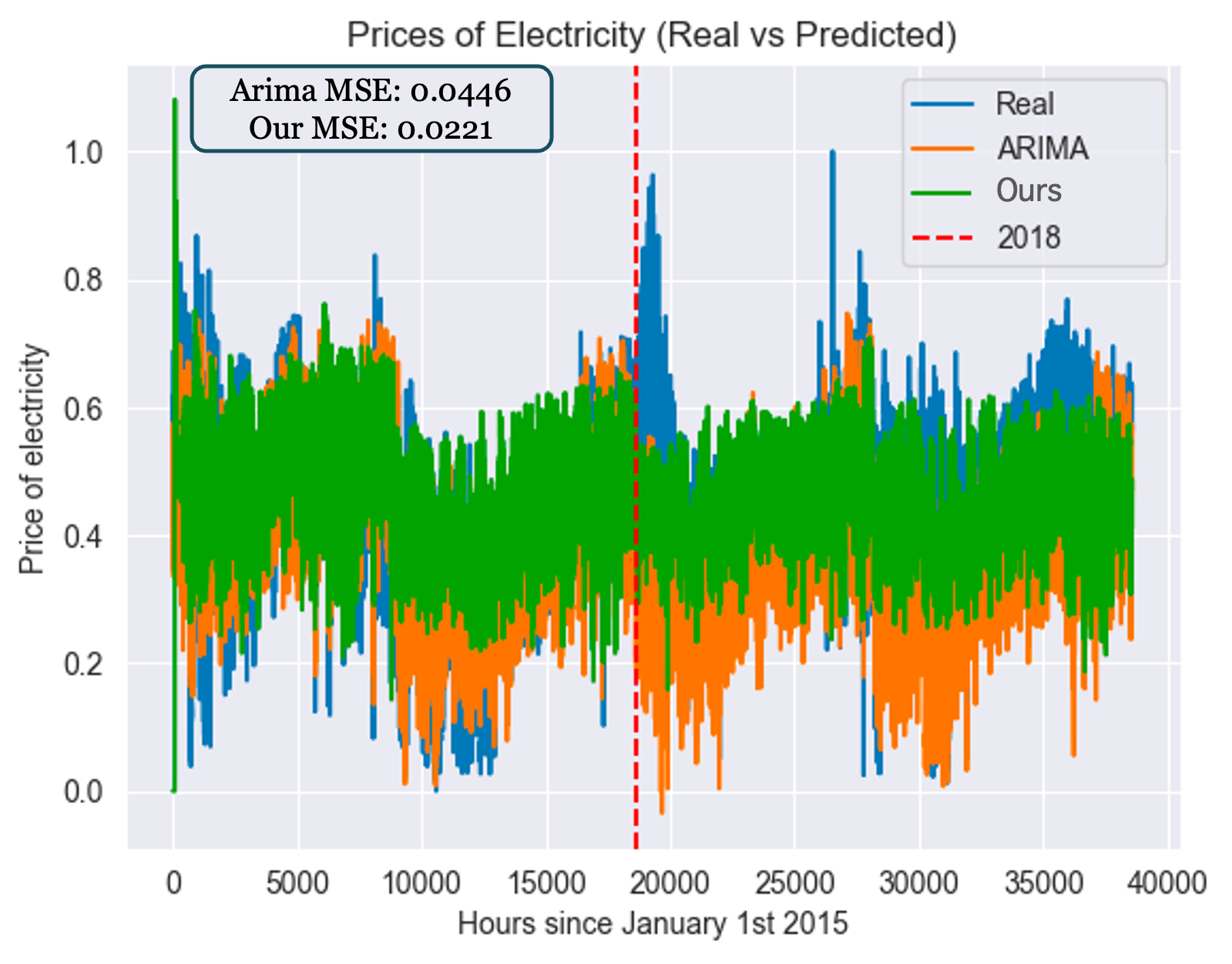}
    \caption{Hourly prices in the Spanish electricity market from January 2015 to December 2020. The Nash simulacrum achieves a MSE that is twice as low as that of the ARIMA method.}
    \label{fig:electricity_market}
    \vspace{-0.5cm}
\end{wrapfigure}

In our first set of experiments, we consider five types of economic games whose equilibria and payoffs have different properties. 
The first three are Fisher market (FM) games, which are zero-sum, between sellers and buyers engaged in trading goods.
These games can be categorized based on the buyers' utility functions as linear, Cobb-Douglas, or Leontief \citep{fisher-tatonnement}.
We then consider two general-sum economic games, which model competition between two firms, namely Cournot competition and Bertrand oligopoly.
When budgets are the only parameters we seek to recover, our min-max formulation is convex-concave, because the players' payoffs are concave in their actions, and affine in their budgets, and hence the regret of players is also affine in the players' budgets. 
In addition, in both the Cournot competition and Bertrand oligopoly games, regret is again convex in the parameters of the game. 
Finally, all the games we study are concave, with the exception of the Bertrand oligopoly game, and the equilibria are unique in the Cobb-Douglas FM, Cournot competition, and Bertrand oligopoly games. 
In each experiment, we generate 500 synthetic game instances, for which the true parameters are known, and use \Cref{alg:gda} (which does not rely on this knowledge) to compute an inverse NE for each.
We record whether our algorithm recovers the true parameters of the market and whether it finds an inverse NE (i.e., average exploitability).
We summarize our findings for the FM games in \Cref{table:both}.
We find that our algorithm recovers the true parameters more often when budgets are the only parameters we seek to recover, as opposed to both budgets and types;
but even in non-convex-concave case, our algorithm is still able to approximate inverse NE over 80\% of the time. 
In settings where the equilibria are unique, we recover true parameters most often, while the worst performance is on Leontief FM games, where payoffs are not differentiable. 


\begin{table}[H]
    \centering
\centering
\begin{minipage}{0.6\linewidth}
    \vspace{-1em}
    \resizebox{1\columnwidth}{!}{%
    \begin{tabular}{|l||c|c|c||c|c|c|}\hline
     Game Parameters & \multicolumn{3}{|c||}{Budgets} & \multicolumn{3}{|c|}{Types + Budgets} \\ \hline \hline
       Fisher Market Type
       & Linear
       & Leontief
       & CD
        & Linear
        & Leontief
        & CD \\ 
     \hline
     \% Parameters Recovered 
     & 100\% & 36.8\% & 100\% 
     & 12\% & 1\% & 99.6\% \\
     \hline
      Average Exploitability 
     & 0.0018 & 0.2240 & 0.0004
     & 0.0119 & 0.1949 & 0.0004 \\
     \hline
     \end{tabular}
     }
\end{minipage}
\begin{minipage}{0.35\linewidth}
\resizebox{1\columnwidth}{!}{%
\begin{tabular}{ |l||c|c|  }
\hline
& Cournot &Bertrand \\
\hline \hline
\% Parameters Recovered
&95.2\% &78\%\\
\hline
Average Exploitability
& 0.0000 & 0.0011\\
\hline
\end{tabular}
}
\end{minipage}

    \caption{The percentage of games for which we recovered the true parameters and the average exploitabilities of the observed equilibrium evaluated w.r.t\@ the computed inverse Nash equilibrium.}
    \label{table:both}
    \vspace{-1em}
\end{table}

In our second set of experiments, we model the Spanish electricity market as a stochastic Fisher market game between electricity re-sellers
and consumers.
In this game, the state comprises the supply of each good and the consumers' budgets, while the re-sellers' actions are to set prices in today's spot market and tomorrow's day ahead market, and the consumers' actions are their electricity demands.
We assume the consumers utilities are linear; this choice is suited to modeling the substitution effect between electricity today and electricity tomorrow.
Using publicly available hourly Spanish electricity prices and aggregate demand data from \href{https://www.kaggle.com/datasets/nicholasjhana/energy-consumption-generation-prices-and-weather}{Kaggle}, we compute a simulacrum of the game that seeks to replicate these observations
from January 2015 to December 2016. 
We also train an ARIMA model on the same data, and run a hyperparameter search for both algorithms using data from January 2017 to December 2018. 
After picking hyperparameters, we then retrain both models on the data between January 2015 to December 2018, and predict prices up to December 2018. 
We also compute the mean squared error (MSE) of both methods using January 2018 to December 2020 as a test set.
We show the predictions of both methods in \Cref{fig:electricity_market}. 
To summarize, we find that the simulacrum makes predictions whose MSE is twice as low.

\bibliography{references}
\bibliographystyle{iclr2024_conference}
\newpage

\section{Appendix}
\section*{Disclaimer} This paper was prepared for informational purposes in part by the Artificial Intelligence Research group of JP Morgan Chase \& Co and its affiliates (``JP Morgan''), and is not a product of the Research Department of JP Morgan. JP Morgan makes no representation and warranty whatsoever and disclaims all liability, for the completeness, accuracy or reliability of the information contained herein. This document is not intended as investment research or investment advice, or a recommendation, offer or solicitation for the purchase or sale of any security, financial instrument, financial product or service, or to be used in any way for evaluating the merits of participating in any transaction, and shall not constitute a solicitation under any jurisdiction or to any person, if such solicitation under such jurisdiction or to such person would be unlawful.

\subsection{Additional Preliminary Definitions}\label{sec_app:prelims}

\paragraph{Notation}
We use caligraphic uppercase letters to denote sets (e.g., $\calX$), bold uppercase letters to denote matrices (e.g., $\allocation$), bold lowercase letters to denote vectors (e.g., $\price$), lowercase letters to denote scalar quantities (e.g., $x$), and uppercase letters to denote random variables (e.g., $X$).
We denote the $i$th row vector of a matrix (e.g., $\allocation$) by the corresponding bold lowercase letter with subscript $i$ (e.g., $\allocation[\buyer])$ and the $j$th entry of a vector (e.g., $\price$ or $\allocation[\buyer]$) by the corresponding Roman lowercase letter with subscript $j$ (e.g., $\price[\good]$ or $\allocation[\buyer][\good]$).
We denote functions by a letter determined by the value of the function: e.g., $f$ if the mapping is scalar-valued, $\f$ if the mapping is vector-valued, and $\calF$ if the mapping is set-valued. We denote the set of natural numbers by $\N$ and the set of real numbers by $\R$.
We denote the positive and strictly positive elements of a set by a $+$ and $++$ subscript, respectively, e.g., $\R_+$ and $\R_{++}$.
For any set $\calC$, we denote its diameter $\max_{\c, \c^\prime \in \calC} \|\c - \c^\prime \|$ by $\diam(\calC)$.
\subsection{Ommited Proofs}\label{sec:app_proofs}

\thminverseNE*
\samy{}
\begin{proof}[Proof of \Cref{thm:inverse_NE}]

By the definition of $\obj$, for all parameter profiles $\param \in \params$,
    \begin{align}
        \max_{\otherstrat \in \stratspace} \obj (\param, \otherstrat) 
        &= \max_{\otherstrat \in \stratspace} \sum_{\player \in \players} \left[\util[\player] (\otherstrat[\player], \truestrat; \param) - \util[\player] (\truestrat; \param) \right] \\
        &= \sum_{\player \in \players} \left[ \underbrace{\max_{\otherstrat[\player] \in \stratspace[\player]} \util[\player] (\otherstrat[\player], \truestrat[-\player]; \param) - \util[\player] (\truestrat; \param)}_{\geq 0} \right] \\
        &\ge 0
    \end{align}

But since the set of inverse NE is non-empty by assumption, $\min_{\param \in \params} \max_{\otherstrat \in \stratspace} \obj (\param, \otherstrat) = 0$. 

But then if $\param^* \in \params$ minimizes $\max_{\otherstrat \in \stratspace} \obj (\param, \otherstrat)$, then $\max_{\otherstrat[\player] \in \stratspace[\player]} \util[\player] (\otherstrat[\player], \truestrat[-\player]; \param^*) - \util[\player] (\truestrat; \param^*) = 0$, for all players $\player \in \players$, which implies $\truestrat$ is a NE and $\param^*$ is an inverse NE.
\end{proof}

\amy{Deni's original proof:}
\begin{proof}[Proof of \Cref{thm:inverse_NE}]
    Notice that for all action profiles $\strat \in \stratspace$ and parameter profile $\param \in \params$, we have:
    \begin{align}
        \max_{\otherstrat \in \stratspace} \obj(\param, \otherstrat) &= \max_{\otherstrat \in \stratspace} \sum_{\player \in \players} \left[\util[\player](\otherstrat[\player], \truestrat; \param) - \util[\player](\truestrat; \param) \right] \\
        &= \sum_{\player \in \players} \left[\max_{\otherstrat[\player] \in \stratspace[\player]} \util[\player](\otherstrat[\player], \truestrat[-\player]; \param) - \util[\player](\truestrat; \param) \right]\\
        &\geq 0
    \end{align}
 
    Additionally, note that under our assumption the set of inverse-NE is non-empty. Hence, there exists a parameter profile $\widehat{\param} \in \params$ such that $\widehat{\param}$ is an inverse-NE,  of $\game[][-1]$, i.e., for all players $\player \in \player$:
    \begin{align}
        \max_{\otherstrat[\player] \in \stratspace[\player]} \util[\player](\otherstrat[\player], \truestrat[-\player]; \widehat{\param}) - \util[\player](\truestrat; \widehat{\param}) = 0
    \end{align}
Summing the above equality across all players, we then have:
\begin{align}
    \sum_{\player \in \players} \max_{\otherstrat[\player] \in \stratspace[\player]} \util[\player](\otherstrat[\player], \truestrat[-\player]; \widehat{\param}) - \util[\player](\truestrat; \widehat{\param}) = 0 \label{eq:min_eq_zero_obj_cumul}
\end{align}


    This means that the minimum of $\max_{\otherstrat \in \stratspace} \obj(\param, \otherstrat)$ is achieved at 0, since for all $\param \in \params$, $\max_{\otherstrat \in \stratspace} \obj(\param, \otherstrat) \geq 0$.

Let $(\param[][][*], \strat[][][][*])$ be any optimal solution to 
$\min_{\param \in \params} \max_{\otherstrat \in \stratspace} \obj (\param, \otherstrat)$.
We will show that $\param[][][*]$ is an inverse NE of $\game[][-1]$.

Since the minimum of $\max_{\otherstrat \in \stratspace} \obj (\param, \otherstrat)$ is achieved at $0$, it follows that
    \begin{align}
        \max_{\otherstrat \in \stratspace} \obj (\param[][][*], \otherstrat) = \sum_{\player \in \players} \underbrace{\max_{\otherstrat[\player] \in \stratspace[\player]} \util[\player] (\otherstrat[\player], \truestrat[-\player]; \param[][][*]) - \util[\player] (\truestrat; \param[][][*])}_{\geq 0} = 0
    \end{align}
But then, since for all players $\player \in \players$, $\max_{\otherstrat[\player] \in \stratspace[\player]} \util[\player] (\otherstrat[\player], \truestrat[-\player]; \param[][][*]) - \util[\player] (\truestrat; \param[][][*]) \geq 0$, it must hold that:
    \begin{align}
        \max_{\otherstrat[\player] \in \stratspace[\player]} \util[\player] (\otherstrat[\player], \truestrat[-\player]; \param[][][*]) - \util[\player] (\truestrat; \param[][][*]) = 0 \\
        \max_{\otherstrat[\player] \in \stratspace[\player]} \util[\player] (\otherstrat[\player], \truestrat[-\player]; \param[][][*]) = \util[\player] (\truestrat; \param[][][*])
    \end{align}
\noindent hence proving that $\truestrat$ is a Nash equilibrium under parameters $\param[][][*]$, i.e., $\param[][][*]$ is an inverse Nash equilibrium.
\end{proof}
\vspace{2em}

\begin{reptheorem}[\ref{thm:concave_game_inverse_NE}]
    Suppose that Assumptions \ref{assum:concave_game}--\ref{assum:smoothness} hold. 
    If \Cref{alg:gda} is run with inputs that satisfy for all $\varepsilon \geq 0$, $\iter \in [\numiters]$ $\learnrate[\otherstrat][\iter] = \learnrate[\param][\iter] = \frac{2\sum_{\player \in \players} \lipschitz[{\grad \util[\player]}]}{\iter}$, and $\numiters \geq \frac{\diam(\params \times \stratspace)}{\varepsilon^2}$ for $\varepsilon \geq 0$, then the time-average of all parameters $\mean[{\param[][\numiters]}] \doteq \frac{1}{\numiters + 1}\sum_{\iter = 0}^\numiters \param[][\iter]$ is an $\varepsilon$-inverse NE, i.e., $\cumulregret[] (\truestrat, \otherstrat; \mean[{\param[][\numiters]}]) - \min_{\substack{\param \in \params}} \max_{\otherstrat \in \stratspace} \cumulregret[] (\truestrat, \otherstrat; \param) \leq \varepsilon$.
\end{reptheorem}

\begin{proof}[Proof of \Cref{thm:concave_game_inverse_NE}]
    The theorem is a direct consequence of Result 3.1 of \citet{nemirovski2009robust}.
\end{proof}
\vspace{1em}
\begin{reptheorem}[\ref{thm:online_sgda}]
    Under \Cref{assum:smooth_convex_invex}, if \Cref{alg:online-sgda} is run with inputs that satisfy for all $\iter \in \numiters$, $\varepsilon \in (0,1)$,  $\learnrate[\otherstrat][\iter] \asymp  \frac{\epsilon^4 \left(\nicefrac{\left\|\nicefrac{\partial\statedist[\initstates][{\policy[][][*]}]}{\partial \initstates} \right\|_\infty}{(1-\discount)}\ssadie{}{\cdot \scparam}\right)^2}{\lipschitz[\grad \obj]^3\left(\lipschitz[\grad \obj]^2 + \sigma^2\right) \left( \nicefrac{\lipschitz[\obj]}{\lipschitz[\grad \obj]} + 1\right)} $,  
    $\learnrate[\param][\iter] \asymp \frac{\varepsilon^8 \left(\nicefrac{\left\|\nicefrac{\partial\statedist[\initstates][{\policy[][][*]}]}{\partial \initstates} \right\|_\infty}{(1-\discount)}
    \ssadie{}{\cdot \scparam}\right)^4}{\lipschitz[\grad \obj]^5 \lipschitz[\obj] \left(\frac{\lipschitz[\obj]}{\lipschitz[\grad \obj]}^2 + 1\right)^4 \left(\lipschitz[\obj]^2 + \sigma^2\right)^{\nicefrac{3}{2}}} \land \frac{\varepsilon^2}{\lipschitz[\grad \obj] \left(\lipschitz[\obj]^2 + \sigma^2\right)}$, and $\numiters \geq \ssadie{\left(\nicefrac{(1-\discount)}{\left\|\nicefrac{\partial\statedist[\initstates][{\policy[][][*]}]}{\partial \initstates} \right\|_\infty} + \frac{\lipschitz[\obj]}{2 \lipschitz[{\grad \obj}]} \right)^{-1} }{\left( 1 + \frac{\lipschitz[\obj]}{2 \lipschitz[{\grad \obj}]} \right)^{-1}}
    \ssadie{\frac{\lipschitz[\obj] \diam(\stratspace \times\params)}{\varepsilon^{2}\learnrate[\otherstrat][\iter]}}{\frac{\lipschitz[\obj] \diam(\stratspace \times\params)}{\varepsilon^{2}\learnrate[\param][\iter]}}$, 
    \sadie{Here the first part should be the 1/gradient dominance param w.r.t $\param$ instead of $\otherstrat$ }
     then the time-average of all parameters $\mean[{\param[][\numiters]}] \doteq \frac{1}{\numiters + 1}\sum_{\iter = 0}^\numiters \param[][\iter]$ is a $\varepsilon$-inverse NE, i.e., $\max_{\strat \in \stratspace} \cumulregret[] (\truestrat, \strat; \mean[{\param[][\numiters]}]) - \min_{\substack{\param \in \params}} \max_{\strat \in \stratspace} \cumulregret[] (\truestrat, \strat; \param) \leq \varepsilon$. 
\end{reptheorem}
\begin{proof}[Proof of \Cref{thm:online_sgda}]
    Firstly, note that under \Cref{assum:convex_param_stoch}, $\obj (\param, \strat)$ is $1$-gradient-dominated in $\param$ since it is \ssadie{concave}{convex} in $\param$ for all $\otherstrat \in \stratspace$ (see Definition 2 of \citet{bhandari2019global}). 

    Now, define  the \mydef{equilibrium distribution mismatch coefficient} $\|\nicefrac{\partial\statedist[\initstates][{\policy[][][\dagger]}]}{\partial \initstates} \|_\infty$ as the Radon-Nikodym derivative of the state-visitation distribution of the Nash equilibrium $\policy[][][\dagger]$ w.r.t.\@ the initial state distribution $\initstates$.
    Under \Cref{assum:smooth_convex_invex}, by \ssadie{Corollary 1}{Theorem 2} and Theorem 4 of \citet{bhandari2019global}, we also have that $\obj (\param, \strat)$ is \ssadie{$\left(\nicefrac{\|\nicefrac{\partial\statedist[\initstates][{\policy[][][\dagger]}]}{\partial \initstates} \|_\infty}{(1- \discount)}\right)$}{$\left(\nicefrac{\|\nicefrac{\partial\statedist[\initstates][{\policy[][][\dagger]}]}{\partial \initstates} \|_\infty}{(1- \discount)} \cdot \scparam\right)$}-gradient-dominated in $\strat$ for all $\param \in \params$.

    \ssadie{}{Moreover, according to analysis in \Cref{sec_app:gradient_estimate}, the variance of the gradient estimator is bounded.}
    Hence, under our Theorem's assumptions, the assumptions of Theorem 2 of \citet{daskalakis2020independent} are satisfied, and we have:
    \begin{align}
        \frac{1}{\numiters + 1} \sum_{\iter = 0}^\numiters \max_{\strat \in \stratspace} \cumulregret[] (\truestrat, \strat; \param[][\iter]) - \min_{\substack{\param \in \params}} \max_{\strat \in \stratspace} \cumulregret[] (\truestrat, \strat; \param) \leq \varepsilon
    \end{align}
    Note that since $\cumulregret[] (\truestrat, \strat, \param)$ is convex in $\param$ for all $\strat \in \stratspace$, then $\strat \mapsto \max_{\strat \in \stratspace} \cumulregret[] (\truestrat, \strat; \param[][\iter])$ is convex by Dankin's theorem \cite{danskin1966thm}. Hence, using convexity, we obtain the theorem's result:
    \begin{align}
         \max_{\strat \in \stratspace} \cumulregret[] (\truestrat, \strat; \frac{1}{\numiters + 1} \sum_{\iter = 0}^\numiters \param[][\iter]) - \min_{\substack{\param \in \params}} \max_{\strat \in \stratspace} \cumulregret[] (\truestrat, \strat; \param) \leq \varepsilon
    \end{align}
    
\end{proof}

\Cref{thm:online_sgda} tells us that in inverse \sdeni{stochastic}{Markov game}s satisfying \Cref{assum:smooth_convex_invex}, an $\varepsilon$-inverse NE can be computed in $\numiters \asymp \nicefrac{\variance^2}{ \varepsilon^{10}} \|\nicefrac{\partial\statedist[\initstates][{\policy[][][*]}]}{\partial \initstates} \|_\infty$.
One way to interpret this result is that the closer the initial state distribution is to the equilibrium state visitation distribution, i.e., the smaller $\left\|\nicefrac{\partial\statedist[\initstates][{\policy[][][*]}]}{\partial \initstates} \right\|$ is, and the smaller the variance $\variance^2$ of the gradient estimators is, the faster the convergence.
On the other hand, if any state that is visited with strictly positive probability by the Nash equilibrium policy is not part of the support of the initial state distribution, then $\left\|\nicefrac{\partial\statedist[\initstates][{\policy[][][*]}]}{\partial \initstates} \right\| \to \infty$, and the convergence bound degrades arbitrarily.

\thminversesimulacrum*

\begin{proof}[Proof of \Cref{thm:inverse_simulacrum}]
    Fix $\mixparamone, \mixparamtwo>0$. 
    Let $(\param[][][*], \policy[][][*])$ be the optimal solutions to the above optimization problem. Notice that for all policy profiles $\policy \in \policies$ and parameter profiles $\param \in \params$, we have $\mixparamone\left\|\policy -  \truepolicy \right\|^2_2 \geq 0$ by the definition of the euclidean norm. Additionally, we have for all $\policy \in \policies$, $\param \in \params$:
    \begin{align}
        \max_{\otherpolicy \in \policies} \mixparamtwo \cumulregret[] (\policy, \otherpolicy; \param) &= \max_{\otherpolicy \in \policies} \mixparamtwo \sum_{\player \in \players} \left[\util[\player] (\otherpolicy[\player], \policy; \param) - \util[\player] (\policy; \param) \right] \\
        &= \mixparamtwo \sum_{\player \in \players} \left[\max_{\otherpolicy[\player] \in \policies[\player]} \util[\player] (\otherpolicy[\player], \policy[-\player]; \param) - \util[\player] (\policy; \param) \right]\\
        &\geq 0
    \end{align}
    Hence, we have: 
    \begin{align}
        \max_{\otherpolicy \in \policies} \otherobj(\param, \policy, \otherpolicy) 
        &= 
        \max_{\otherpolicy \in \policies} \left\{ \mixparamone \Ex_{\substack{(\obs,\trueobs) \sim \obsdistrib[][][\policy] \times \obsdistrib[][][\truepolicy]}} \left[\left\|\obs - \trueobs \right\|^2 \right] + \mixparamtwo \cumulregret[] (\policy, \otherpolicy; \param) \right\}\\
        &=  \mixparamone \Ex_{\substack{(\obs,\trueobs) \sim \obsdistrib[][][\policy] \times \obsdistrib[][][\truepolicy]}} \left[\left\|\obs - \trueobs \right\|^2 \right] + \max_{\otherpolicy \in \policies} \mixparamtwo \cumulregret[] (\policy, \otherpolicy; \param)\\ 
        &\geq \mixparamone (0) + \mixparamtwo (0) = 0
    \end{align}
    Additionally, note that under our assumption the set of inverse-NE is non-empty. Hence, there exists a tuple of parameter and action profiles $(\param[][][*], \policy[][][*])$ such that $\policy[][][*] = \truepolicy$, and $\param[][][*]$ is an inverse-NE,  of $(\numplayers, \numactions, \policies, \params, \util, \truepolicy)$:
%
%
    \begin{align}
        &\max_{\otherpolicy \in \policies} \otherobj(\param[][][*], \policy[][][*], \otherpolicy) \\
        &= \max_{\otherpolicy \in \policies} \left\{\mixparamone\Ex_{\substack{(\obs,\trueobs) \sim \obsdistrib[][][{\policy[][][*]}] \times \obsdistrib[][][\truepolicy]}}  \left[\left\|\obs - \trueobs \right\|^2 \right] + \mixparamtwo \cumulregret[] (\truepolicy, \otherpolicy; \param[][][*]) \right\}\\
        &=  \mixparamone\Ex_{\substack{(\obs,\trueobs) \sim \obsdistrib[][][{\policy[][][*]}] \times \obsdistrib[][][\truepolicy]}}  \left[\left\|\obs - \trueobs \right\|^2 \right] + \mixparamtwo \max_{\otherpolicy \in \policies} \cumulregret[] (\truepolicy, \otherpolicy; \param[][][*])\\ 
        &= \mixparamone (0) + \mixparamtwo (0) = 0  \label{eq:min_eq_zero_obj}
    \end{align}

\noindent where the final line follows from the definition of the inverse Nash equilibrium, i.e., $\cumulregret[] (\truepolicy, \otherpolicy; \param[][][*]) = 0$.

    This in turn means that the minimum of $\max_{\otherpolicy \in \policies} \otherobj(\param, \policy, \otherpolicy)$ is achieved at 0. 
    
    Finally, we show that any tuple $(\param[][][*], \policy[][][*])$ of parameter and action profiles which are a minimum of $\max_{\otherpolicy \in \policies} \otherobj(\param, \policy, \otherpolicy)$, i.e. $(\param[][][*], \policy[][][*]) \in \argmin_{\substack{\param \in \params \\ \policy \in \policies}} \max_{\otherpolicy \in \policies} \otherobj(\param, \policy, \otherpolicy)$ respectively correspond to a tuple $(\param[][][*], \policy[][][*])$ such that $\Ex_{\substack{(\obs,\trueobs) \sim \obsdistrib[][][{\policy[][][*]}] \times \obsdistrib[][][\truepolicy]}}$, and $\param[][][*]$ is an inverse-NE of $(\numplayers, \numactions, \policies, \params, \util, \policy[][][*])$.

    Recall that, by \Cref{eq:min_eq_zero_obj}, we have: 
    \begin{align}
        \min_{\substack{\param \in \params \\ \policy \in \policies}} \max_{\otherpolicy \in \policies} \otherobj(\param, \policy, \otherpolicy) &= \min_{\substack{\param \in \params \\ \policy \in \policies}} 
        \max_{\otherpolicy \in \policies} \left\{ \mixparamone\left\|\truepolicy - \policy \right\|^2_2 + \mixparamtwo \cumulregret[] (\policy, \otherpolicy; \param) \right\}\\ &= \min_{\substack{\param \in \params \\ \policy \in \policies}} 
        \left\{ \mixparamone \underbrace{\left\|\truepolicy - \policy \right\|^2_2}_{\geq 0} + \underbrace{\max_{\otherpolicy \in \policies} \mixparamtwo \cumulregret[] (\policy, \otherpolicy; \param)}_{\geq 0} \right\} \\
        &= 0 
    \end{align}

    Hence, it must be that $\left\|\truepolicy - \policy[][][*] \right\|^2_2 = 0$ and $\max_{\otherpolicy \in \policies} \cumulregret[] (\policy[][][*], \otherpolicy; \param[][][*]) = 0$, proving the desired result.
\end{proof}

    

As the computation of a simulacrum is in general a non-convex-non-concave problem, we cannot compute a solution to the min-max optimization in polynomial-time, however, we can obtain best iterate convergence to a stationary point of the \mydef{Moreau envelope of the empirical exploitability} $\regulexploit(\param, \strat) \doteq \min_{(\param[][][\prime], \strat[][][][\prime]) \in \params \times \stratspace} \left\{ \exploit(\param, \strat) + \lipschitz[{\grad \avg[\otherobj]}]\left\| (\param, \strat) - (\param[][][\prime], \strat[][][][\prime])\right\|^2\right\}$, i.e., a point $(\param, \strat) \in \params \times \stratspace$ \sadie{This exploitability is not empirical here} s.t. $\left\|\grad \regulexploit(\param, \strat) \right\| = 0$ under suitable assumptions satisfied by a large class of Markov games including discrete state and action space Markov games.%
\footnote{We note that stationary points of the Moreau envelope correspond to stationary points of the subgradient of $\avg[\exploit]$, but as exploitability \amy{exploitability!!! uh oh!!!} is not neccessarily differentiable, the Moreau envelope is used to measure distance to a stationary point~\cite{lin2020gradient}.}

\begin{reptheorem}[\ref{thm:apprenticeship_thm}]\label{thm_app:apprenticeship_thm}
    Suppose that \Cref{assum:smooth_convex_invex} holds, and assume in addition that for all $\policy[][][\strat] \in \policies[][\stratspace]$ , $\obsdistrib[][][{\policy[][][\strat]}]$ is twice continuously differentiable in $\strat$.  
    If \Cref{alg:offline-sgda} is run with inputs that satisfy for all $\iter \in [\numiters]$ $\varepsilon \in (0,1)$,  $\learnrate[\otherstrat][\iter] \asymp  \frac{\epsilon^4 \left(\nicefrac{\left\|\nicefrac{\partial\statedist[\initstates][{\policy[][][*]}]}{\partial \initstates} \right\|_\infty}{1-\discount}\ssadie{}{\cdot \scparam}\right)^2}{\lipschitz[\grad \obj]^3\left(\lipschitz[\grad \obj]^2 + \sigma^2\right) \left( \nicefrac{\lipschitz[\obj]}{\lipschitz[\grad \obj]} + 1\right)} $,  
    $\learnrate[\param][\iter] \asymp \frac{\varepsilon^8 \left(\nicefrac{\left\|\nicefrac{\partial\statedist[\initstates][{\policy[][][*]}]}{\partial \initstates} \right\|_\infty}{1-\discount} \ssadie{}{\cdot \scparam}\right)^4}{\lipschitz[\grad \obj]^5 \lipschitz[\obj] \left(\frac{\lipschitz[\obj]}{\lipschitz[\grad \obj]}^2 + 1\right)^4 \left(\lipschitz[\obj]^2 + \sigma^2\right)^{\nicefrac{3}{2}}} \land \frac{\varepsilon^2}{\lipschitz[\grad \obj] \left(\lipschitz[\obj]^2 + \sigma^2\right)}$, \sadie{Do we have bound on $\numiters$ here?}
      then the best-iterate parameters and policies $(\bestiter[{\param}][\numiters], \bestiter[{\strat}][\numiters]) \in \argmin_{\iter \in [\numiters]} \left\|\grad \regulexploit(\param[][\iter], \strat[][][\iter]) \right\|$ converge to a stationary point of the exploitability, i.e., $\left\|\grad \regulexploit(\bestiter[{\param}][\numiters], \bestiter[{\strat}][\numiters]) \right\| \leq \varepsilon$. 

     Additionally, for any $\zeta, \xi \geq 0$ and for a sample size of equilibrium observations $\numsamples \asymp \nicefrac{1}{\xi^2} \log(\nicefrac{1}{\zeta})$, with probability $1-\zeta$, we have:
        \begin{align}
            \avg[{\exploit}] (\bestiter[{\param}][\numiters], \bestiter[{\strat}][\numiters]) - \exploit(\bestiter[{\param}][\numiters], \bestiter[{\strat}][\numiters]) \leq \xi
        \end{align}
\end{reptheorem}

\begin{proof}[Proof of \Cref{thm_app:apprenticeship_thm}]
    Although \cite{daskalakis2020independent}'s Theorem 2 is stated for functions which are gradient-dominated-gradient-dominated, their proof falls through for any function which is non-convex-gradient-dominated. Define  the \mydef{equilibrium distribution mismatch coefficient} $\|\nicefrac{\partial\statedist[\initstates][{\policy[][][\dagger]}]}{\partial \initstates} \|_\infty$ as the Radon-Nikodym derivative of the state-visitation distribution of the Nash equilibrium $\policy[][][\dagger]$ w.r.t.\@ the initial state distribution $\initstates$.
    Under \Cref{assum:smooth_convex_invex}, by Corollary 1 and Theorem 4 of \citet{bhandari2019global}, we have that $\otherobj(\param, \strat, \otherstrat)$ is\ssadie{ $\left(\nicefrac{\|\nicefrac{\partial\statedist[\initstates][{\policy[][][\dagger]}]}{\partial \initstates} \|_\infty}{(1- \discount)}\right)$}{ $\left(\nicefrac{\|\nicefrac{\partial\statedist[\initstates][{\policy[][][\dagger]}]}{\partial \initstates} \|_\infty}{(1- \discount)}\cdot \scparam\right)$}-gradient-dominated in $\otherstrat$ for all $\param \in \params$ and $\strat \in \stratspace$.

    
       \ssadie{}{Moreover, according to analysis in \Cref{sec_app:gradient_estimate}, the variance of the gradient estimator is bounded.} Hence, under our Theorem's assumptions, the assumptions of Theorem 2 of \citet{daskalakis2020independent} are satisfied, and we have:
    \begin{align}
        \frac{1}{\numiters + 1} \sum_{\iter = 0}^\numiters \left\|\grad \regulexploit(\param[][\iter], \strat[][][\iter]) \right\| \leq \varepsilon
    \end{align}
Taking a minimum across all $\iter = 0, 1, \hdots, \numiters$, we then have:
    \begin{align}
          \min_{\iter = 0, 1, \hdots, \numiters} \left\|\grad \regulexploit(\param[][\iter], \strat[][][\iter]) \right\| \leq \varepsilon
    \end{align}

    The second part is then a direct consequence of the hoeffding bound, whose assumptions are satisfied since the objective is bounded from above and from below by 0, as the objective is continuous and its domain is non-empty, and compact.
\end{proof}

\newpage
\subsection{Related Work}

\paragraph{Microeconomics}

The literature on characterizing agent preferences that can be rationalized by payoff functions, known under the names of \mydef{reveled preference theory} \cite{samuelson1948consumption, afriat1967construction, varian1982nonparametric, varian2006revealed} and the \mydef{integrability problem} \cite{mas-colell}, far predates concerns of computing payoff functions that generate observed behavior.
While revealed preference theory is concerned with understanding when a set of observed purchasing decisions for a consumer and associated market conditions (e.g., prices) is consistent with payoff-maximizing behavior, the integrability problem aims to characterize those consumption functions that can arise as the solution to a payoff maximization problem.\amy{i can't grok the integrability problem from this description.}\deni{Integrability assumes you observe the whole demand function, while learning from revealed preferences assumes you observe only samples from the demand function.}
The difference between revealed preference theory and the integrability problem is analogous to the difference between inverse optimization and inverse learning.

\paragraph{Econometrics}

A large body of work in the econometrics literature is dedicated to inverse game theory\amy{it doesn't make sense to me that the econometrics people would be doing IGT, and not IRL. they are all statistics all the time, and typically infer whatever it is they infer from samples}, with a recent focus on inferring bidders' valuations in online auctions. \citeauthor{nekipelov2015econometrics} \citeyear{nekipelov2015econometrics} analyzed inferring bidders' utilities in online ad auctions, assuming bidders are no-regret learners, and hence learn (coarse) correlated equilibrium.
\citeauthor{syrgkanis2017inference} \citeyear{syrgkanis2017inference} propose a method that infers agent types, assuming they play a Bayes-Nash equilibrium.
More broadly, the identification literature \cite{bresnahan1991empirical, lise2001estimating, bajari2010identification} is closely related to our work, but usually does not address computational complexity concerns.
Furthermore, the settings considered in this literature are overwhelmingly \sdeni{one-shot}{normal-form} Bayesian, game, while our primary focus in this paper is (complete-information) stochastic games.

\paragraph{Inverse Optimization.} 

Inverse optimization \cite{heuberger2004inverse,chan2021inverse} seeks to recover the parameters of an optimization problem given access to the solution of the problem. One of the central results in inverse optimization demonstrates that one can recover the objective function of any linear inverse optimization problem from its solution by solving a linear program \cite{chan2022inverse}.
Our work considers the more general inverse problem for multiple agents with arbitrary objective, i.e., payoff functions, and solves it using a mathematical program as well.

\paragraph{Inverse Algorithmic Game Theory}

A literature that lies at the intersection of economics and computer science has aimed to provide computationally-efficient methods for rationalizing equilibria, but has mainly focused on specific types of games, such as matching \cite{kalyanaraman2008complexisaac} and network formation games \cite{kalyanaraman2009complexfocs}.
In the latter case, they showed that game attributes that are local to a player can be rationalized.
More recently, \citeauthor{kuleshov2015inverse}
(\citeyear{kuleshov2015inverse}) showed that correlated equilibria can be rationalized in polynomial-time in succint games.
We note that these computational results concern stylized game models, and restrict certain aspects of the game, such as the size of the game's parameter space.
In contrast, our results abstract away the issue of efficiently representatng the game's parameter space, and show that under appropriate parameterization,
inverse equilbirium can be computed in polynomial time.

\paragraph{Inverse Reinforcement Learning} 

Algorithms that infer the reward function of an agent operating within a Markov decision process \cite{bellman1952theory} have been studied extensively in recent years, starting with the initial investigations by \citet{ng2000algorithms}.
These algorithms can be broadly categorized as maximum margin methods \cite{ratliff2006maximum, silver2008high, abbeel2004apprenticeship, syed2007game}, i.e., methods that seek to maximize the margin between the value of observed behavior and the behavior associated with learned policy and rewards;
maximum entropy methods \cite{ziebart2008maximum, wulfmeier2015maximum, ziebart2008maximum,theodorou2010generalized, boularias2012structured, boularias2011relative}, i.e., methods that maximize the entropy of the observed and the learned behaviors;
Bayesian learning methods \cite{ramachandran2007bayesian, choi2011map,lopes2009active,levine2011nonlinear,babes2011apprenticeship}, i.e., methods that learn a posterior distribution over parameters using Bayesian updating;
and classification/regression methods \cite{klein2012inverse,taskar2005learning,klein2013cascaded,brown2019extrapolating}, i.e., methods that learn parameters that minimize the distance between the observed behavior and behavior generated by learned behavior using the inferred parameters.
Our methods, when used with only one player, can be characterized as a class of novel \samy{regret-minimizing}{} inverse reinforcement learning methods, which seek to recover parameters that minimize the players' regrets.
\amy{i really want to say minimize exploitability, but i think regret is a better understood term, colloquially, and i am not sure we have defined exploitability in the intro?}

\paragraph{Inverse Game Theory}

In multiagent settings, convex programming formulations have been proposed for inferring game parameters  \amy{what is the diff b/n these two settings? inverse game theory and the inverse eqm problem? i can't make sense of what's written.} \deni{changed word ordering, should be good now!}in normal-form games under the names of the inverse game theory problem \cite{kuleshov2015inverse} and the inverse equilibrium problem \cite{waugh2013computational, bestick2013inverse}.
These methods focus on computing an inverse correlated equilibrium, and in the case of \citeauthor{waugh2013computational}, further seek to reproduce observed equilibrium behavior via a maximum entropy correlated equilibrium.
\citeauthor{hadfield2016cooperative} \citeyear{hadfield2016cooperative} consider cooperative inverse reinforcement learning, which can be seen as an inverse Nash equilibrium problem in a particular zero-sum imperfect-information game, but this method is not accompanied by computational guarantees.

\paragraph{Multiagent Inverse Reinforcement Learning}

Multiagent inverse reinforcement learning generalizes inverse game theory from normal-form games to Markov games \cite{natarajan2010multi, lin2019multi, yu2019multi, lin2017multiagent, fu2021evaluating}.
Even more importantly, instead of observing equilibrium policies, only sample trajectories from equilibrium policies are observed.
\citet{lin2019multi} study the inverse Nash equilibrium problem in zero-sum games \cite{lin2017multiagent}, and extend their methods to solve for inverse correlated equilbrium in general-sum stochastic games, and inverse Nash equilibrium in a restricted class of 
adversarial stochastic games.
\citeauthor{yu2019multi} (\citeyear{yu2019multi}) propose gradient-based algorithms for computing inverse quantal response equilibria with function approximation.

\newpage
\subsection{Markets Experiments}\label{sec_app:markets}

\subsubsection{Static Fisher Markets} 
A \mydef{(one-shot) Fisher market} consists of $\numbuyers$ buyers and $\numgoods$ divisible goods with unit supply\citep{brainard2000compute}.
Each buyer $\buyer \in \buyers$ is endowed with a budget $\budget[\buyer] \in \budgetspace[\buyer] \subseteq \mathbb{R}_{+}$ and a utility function $\util[\buyer]: \mathbb{R}_{+}^{\numgoods} \times \typespace[\buyer] \to \mathbb{R}$, which is parameterized by a type $\type[\buyer] \in \typespace[\buyer]$ that defines a preference relation over the consumption space $\R^\numgoods_+$. An instance of a Fisher market is then a tuple $\calM \doteq (\numbuyers, \numgoods, \util, \type, \budget)$, where $\util \doteq \left(\util[1], \hdots, \util[\numbuyers] \right)$ is a vector-valued function of all utility functions and $\budget \doteq (\budget[1], \hdots, \budget[\numbuyers]) \in \R_{+}^{\numbuyers}$ is the vector of buyer budgets.
When clear from context, we simply denote $\calM$
by $(\type, \budget)$.

Given a Fisher market $(\type, \budget)$, an \mydef{allocation} $\allocation = \left(\allocation[1], \hdots, \allocation[\numbuyers] \right)^T \in \R_+^{\numbuyers \times \numgoods}$ is a map from goods to buyers, represented as a matrix, s.t. $\allocation[\buyer][\good] \ge 0$ denotes the amount of good $\good \in \goods$ allocated to buyer $\buyer \in \buyers$. Goods are assigned \mydef{prices} $\price = \left(\price[1], \hdots, \price[\numgoods] \right)^T \in \mathbb{R}_+^{\numgoods}$. A tuple $(\allocation[][][][*], \price^*)$ is said to be a \mydef{competitive equilibrium (CE)} \citep{arrow-debreu, walras} if 
1.~buyers are utility maximizing, constrained by their budget, i.e., $\forall \buyer \in \buyers, \allocation[\buyer][][][*] \in \argmax_{\allocation[ ] : \allocation[ ] \cdot \price^* \leq \budget[\buyer]} \util[\buyer](\allocation[ ], \type[\buyer])$;
and 2.~the market clears, i.e., $\forall \good \in \goods,  \price[\good]^* > 0 \Rightarrow \sum_{\buyer \in \buyers} \allocation[\buyer][\good][][*] = \supply[\good]$ and $\price[\good]^* = 0 \Rightarrow\sum_{\buyer \in \buyers} \allocation[\buyer][\good][][*] \leq \supply[\good]$.

The set of CE of any Fisher market $(\type, \budget)$ with continuous, concave, and homogeneous\footnote{A function $f: \R^m \to \R$ is called \mydef{homogeneous of degree $k$} if $\forall \allocation[ ] \in \R^m, \lambda > 0, f(\lambda \allocation[ ]) = \lambda^k f(\allocation[ ])$.} utility functions is equal to the set of Nash equilibria of the \mydef{Eisenberg-Gale min-max game},%
\footnote{This min-max game corresponds to the Lagrangian saddle-point formulation of the Eisenberg-Gale program \cite{gale1989theory, jain2005market}.} 
a convex-concave min-max game between a seller who chooses prices $\price \in \R_+^{\numgoods}$ and buyers who collectively choose allocations
$\allocation \in \R_+^{\numbuyers \times \numgoods}$:
the objective function of this game comprises two sums: the first is the logarithmic Nash social welfare of the buyers' utility \ssadie{per budget, i.e., bang-per-buck}{}, while the second is the profit of a fictional auctioneer who sells the goods in the market:
\begin{align}
    \min_{\price \in \R_+^{\numgoods}} \max_{\allocation \in \R_+^{\numbuyers \times \numgoods}} 
    & \obj (\price, \allocation; \type, \budget) \doteq \sum_{\buyer \in \buyers} \budget[\buyer] \log \left( \util[\buyer](\allocation[\buyer], \type[\buyer]) \right) + \sum_{\good \in \goods} \left( \price[\good] - \price[\good] 
    \sum_{\buyer \in \buyers}\allocation[\buyer][\good] \right)
    \label{eq:eg_game}
\end{align}

Therefore, for any Fisher market $\calM \doteq (\numbuyers, \numgoods, \util, \typetrue, \budgettrue)$, 
we can construct a inverse game $\game[][-1] \doteq 
(\game[][\paramtrue]/\paramtrue , \truestrat)$ where $\game[][\paramtrue]$ is the corresponding Eisenberg-Gale min-max game (\Cref{eq:eg_game}) parameterized by the true types and budgets $\paramtrue=(\typetrue, \budgettrue)$, and $\truestrat=(\allocation[][][][*], \price^*)$ is not only a NE of the game $\game[][\paramtrue]$ but also a CE of market $\calM$. Our goal is to recover the true market parameters $\paramtrue=(\typetrue, \budgettrue)$ given the observed $\truestrat=(\allocation[][][][*], \price^*)$, by solving this inverse game problem using \Cref{thm:inverse_NE} and \Cref{alg:gda}. 

We ran two different experiments\footnote{We include a detailed description of our experimental setup in the appendix.}. First, we solved a simpler inverse game problem where the true type $\typetrue$ is given, and we just need to retrieve the true budgets $\budgettrue$; then, we attempt to recover both true type and true budgets simultaneously. For both experiments, we created 500 markets with each of these three (standard) classes of utility functions parameterized by types:
1.~\mydef{linear}: $\util[\buyer](\allocation[\buyer]; \type[\buyer]) = \sum_{\good \in \goods} \type[\buyer][\good] \allocation[\buyer][\good]$; 2.~\mydef{Cobb-Douglas (CD)}:  $\util[\buyer](\allocation[\buyer]; \type[\buyer]) = \prod_{\good \in \goods} {\allocation[\buyer][\good]}^{\type[\buyer][\good]}$; and 3.~\mydef{Leontief}:  $\util[\buyer](\allocation[\buyer]; \type[\buyer]) = \min_{\good \in \goods} \left\{ \frac{\allocation[\buyer][\good]}{\type[\buyer][\good]}\right\}$. 
Then, we ran \Cref{alg:gda} on min-max optimization problem \sadie{I may write out the specific min-max problem for fisher, but I don't think we have enough space. Maybe in the Appendix.} defined in \Cref{eq:min_max_gen_sim} to compute the inverse Nash Equilibrium of the inverse game $\game[][-1]$ defined above.
Finally, for each utility type, we recorded the percentage of markets that we could recover the true parameters, i.e., the markets for which our computed parameters is close enough to the true parameters, and the average exploitability\sadie{refer appendix?} of the observed equilibrium evaluated under the computed parameters across all markets.

\Cref{table:both} shows that when only retrieving budgets,we were able to recover all the parameters and minimize exploitability in markets with Linear and Cobb-Douglas utilities, but we hardly do so in Leontief markets. The difficulty in this case likely arises from two aspects: first, Leontief utility function is not differentiable, so the min-max optimization problems associated to Leontief
markets are not smooth; moreover, for any Leontief Fisher market, the CE is not guaranteed to be unique. 
When computing budgets and types at once, while our algorithm can still minimize exploitability for both Linear nad Cobb-Douglas markets, it cannot really retrive true parameters in Linear markets. This may due to the fact that, in Linear markets, though competitive equilibrium prices are unique, the competitive allocations are not guaranteed to be unique; by contrast, the CEs are always unique in Leontief markets. 

\paragraph{Hyperparameters}
We randomly initialized 500 different linear, Leontief, Cobb-Douglas Fisher markets, each with 3 buyers and 2 goods. Buyer $\buyer$’s budget $\budget[\buyer]$ was drawn randomly from a uniform distribution ranging from 0 to 10 (i.e., $U[0,10]$), and each buyer $\buyer$’s type for good $\good$, $\type[\buyer][\good]$, was drawn randomly from $U[0,10]$. 

For Fisher markets with all three class of utilities, we ran our algorithm for 5000 iterations with learning rate $\learnrate[\param]=0.01$. Moreover, we stop our algorithm when our computed inverse Nash equilibrium $\param$ is closed enough to the original parameter $\paramtrue$: $||\frac{\param-\paramtrue}{\paramtrue}||_2\leq \epsilon$ where we set $\epsilon=0.1$.

\subsubsection{Cournot Competition and Bertrand Competition}
A \mydef{Cournot competition model} $\calC \doteq (\numfirms, \mcost, \pricefunc)$ consists of $n$ firms that produce a homogeneous product, and each firm $i$ chooses a quantity level of production $\Cprod[i]$ that maximizes its profits. All firms face a marginal cost $\mcost$. That is, for a given firm $i$, the cost of producing $\Cprod[i]$ unit of good is $\mcost\Cprod[i]$. The price function $\pricefunc$ takes the total production of all firms $Q_{total}=\sum_{i\in \firms}\Cprod[i]$ as input and outputs the unit prices for the good. Thus, the profit function for firm $i$ is $f_i(\Cprod[i], \Cprod[-i]; \mcost)=\Cprod[i](\pricefunc(\sum_{i\in \firms}\Cprod[i])-\mcost)$. $\Cprod^*$ is a Nash equilibria of the Cournot game if and only if $\Cprod[i]^* \in \argmax_{\Cprod[i]in \R_+} f_i(\Cprod[i], \Cprod[-i]^*; \mcost)$ for all $i\in \firms$.

A \mydef{Bertrand competition model} $\calB \doteq (\numfirms, \mcost, \demandfunc)$ is also a competition model that consists of $n$ firms producing a homogeneous product, but this time, each firm $i$ set prices $\Bprices[i]$ to maximize its profits. All firms face a marginal cost $\mcost$. That is, for a given firm $i$, the cost of producing $\Cprod[i]$ unit of good is $\mcost\Cprod[i]$. The demand function $\demandfunc$ takes the minimum price proposed by the firms $\Bprices[\min]=\min_{i\in \firms} \Bprices[i]$ as input and outputs the demand for that good in the whole market. Firm $i$’s individual demand function is a function of the price set by each firm: 
\begin{align}
    D_{i}(\Bprices[i],\Bprices[-i])=\begin{cases}
        D(\Bprices[\min])  & \Bprices[i]=\Bprices[\min], \Bprices[j]\geq \Bprices[\min] \forall j\neq i\\
        \frac{D(\Bprices[\min])}{n} &\Bprices[i]=\Bprices[\min],\text{$n=$\# of $j\in \firms$ with $\Bprices[j]=\Bprices[\min]$}\\
        0 & \Bprices[i]\neq \Bprices[\min]
    \end{cases}
\end{align}


Thus, the profit function for the firm $\i$ is $f_i(\Bprices[i], \Bprices[-i]; \mcost)=D_{i}(\Bprices[i],\Bprices[-i])(\Bprices[i]-\mcost)$. $\Bprices^*$ is a Nash equilibria of the Bertrand game if and only if $\Bprices[i]^* \in \argmax_{\Bprices[i]in \R_+} f_i(\Bprices[i], \Bprices[-i]^*; \mcost)$ for all $i\in \firms$.

In experiments, we generated 500 Cournot competition models and 500 Bertrand competition models and attempted to retrieve their true parameter, i.e., marginal costs, given equilibrium productions/equilibrium prices respectively. \Cref{table:both} shows that our algorithm can effectively recover the true parameters in Cournot games and minimize the exploitability of the observed equilibrium evaluated under the computed parameters. In the Bertrand games, though we could only recover $78\%$ true parameters, the average exploitability of the observed equilibrium evaluated under the computed inverse Nash equilibrium is mostly minimized. 

\paragraph{Hyperparameters}
We randomly initialized 500 different duopoly Cournot competitions and duopoly Bertrand competitions. Marginal costs was drawn randomly from a uniform distribution ranging from 2 to 20 (i.e., $U[2,20]$) for both Cournot and Bertrand. Moreover, we define the price functions in Cournot competitions as $P(\Cprod[total]) = a + b\Cprod[total]$ where $a\sim U[10,100]$ and $b\sim U[-10, -0.01]$. We define the demand functions in Bertrand competitions as $D(\Bprices[\min])=c+d\Bprices[\min]$ where $c\sim U[10,100]$ and $d\sim U[-10, -0.01]$.

For Cournot competitions, we ran our algorithm for 10000 iterations with learning rate $\learnrate[\param]=0.01$, and for Bertrand competitions, we ran our algorithm for 250 iterations with learning rate $\learnrate[\param]=0.3$. Moreover, we stop our algorithm when our computed inverse Nash equilibrium $\param$ is closed enough to the original parameter $\paramtrue$: $||\frac{\param-\paramtrue}{\paramtrue}||_2\leq \epsilon$ where we set $\epsilon=0.1$.

\subsubsection{Stochastic Fisher Markets}
\mydef{A (static) Fisher market} $(\numbuyers, \numgoods, \consumptions, \util, \supply, \type, \budget)$, $(\supply, \type, \budget)$ when clear from context, consists of $\numbuyers \in \N_{++}$ buyers and $\numgoods \in \N_{++}$ divisible goods \cite{brainard2000compute}.
Each buyer $\buyer \in \buyers$ is represented by a tuple $(\consumptions[\buyer], \util[\buyer],\type[\buyer], \budget[\buyer])$, $(\type[\buyer], \budget[\buyer])$ when clear from context, which consists of a \mydef{budget} $\budget[\buyer] \in \budgetspace[\buyer] \subseteq \R_+$ of some num\'eraire good it is endowed with, a \mydef{utility function}
$\util[\buyer]: \consumptions[\buyer] \times \typespace[\buyer] \to \mathbb{R}_+$, which is parameterized by a \mydef{type} $\type[\buyer] \in \typespace[\buyer]$ s.t. $\util[\buyer](\ \cdot \ ; \type[\buyer])$ defines a preference relation over the \mydef{consumption space} $\consumptions[\buyer] \subseteq \R^\numgoods_+$.
Each good is characterized by a supply $\supply[\good] \in \supplyspace[\good] \subseteq \R_+$. We denote the collection of all utility functions $\util \doteq \left(\util[1], \hdots, \util[\numbuyers] \right)$, the collection of buyer types $\type \doteq \left(\type[1], \hdots, \type[\numbuyers]\right)$, the collection of buyer budgets $\budget \doteq (\budget[1], \hdots, \budget[\numbuyers]) \in \R_{+}^{\numbuyers}$, the collection of all good supplies $\supply \doteq (\supply[1], \hdots, \supply[\numgoods]) \in \R_{+}^{\numgoods}$, the joint space of consumptions $\consumptions \doteq \bigtimes_{\buyer \in \buyers} \consumptions[\buyer]$, the joint space of  types $\typespace \doteq \bigtimes_{\buyer \in \buyers} \typespace[\buyer]$, and the joint space of budgets $\budgetspace \doteq \bigtimes_{\buyer \in \buyers} \budgetspace[\buyer]$.


\begin{definition}[Stochastic Fisher Market Game]
    Given a stochastic Fisher market $\fishermkt \doteq (\numbuyers, \numgoods, \numassets, \states, \util, \trans, \discount, \initstates)$, we define the Stochastic Fisher Market Game $\mgame[][\fishermkt] \doteq (\numplayers, \states, \outeractions, \inneractions, \reward, \trans, \discount, \initstates)$ where:
    \begin{alignat}{3}
        &\states \doteq \worldstates \times \supplyspace \times \budgetspace \times \typespace
        \qquad \qquad&\outeractions \doteq \pricespace
        \qquad \qquad&\inneractions(\state) \doteq \consumptions \times \savings(\state)
    \end{alignat}
    \begin{align}
        \reward((\worldstate, \supply, \type, \budget), \price, (\allocation, \saving)) \doteq  \price \cdot \left( \supply - \sum_{\buyer \in \buyers} \allocation[\buyer] \right) + \sum_{\buyer \in \buyers} \left(\budget[\buyer] + \saving[\buyer] \right) \log \left( \frac{\util[\buyer](\allocation[\buyer]; \type[\buyer])}{\budget[\buyer] + \saving[\buyer] }\right)
    \end{align}
\end{definition}


\paragraph{Experimental setup}
We use Jax, and Haiku to traint he simulacrum policies and use a feedforward neural network with 4 layers with 200 nodes. We run our experiments with 5 random seed and report the best results.
\newpage
\subsection{Gradient Estimators}\label{sec_app:gradient_estimate}

Notice that the deterministic policy gradient theorem tells us that to compute a the policy gradient we need to compute the gradient of state-action value function with respect to the actions and then multiply it by the gradient of the policy w.r.t. the policies. Since we have access to a first order oracle of the reward and transition we can then compute gradient of the cumulative regret with the following quantities:

\begin{align}
    &\obj[\strat](\param, \otherstrat; \histmatrix, \hist[][\prime]) \notag \\ 
    &\propto \sum_{\player \in \players} \left[ \grad[\action] \reward[\player](\state[0][][\player,], \action[][][0][\player,]; \param) \right. \notag \\&+ \left. \discount \grad[\action] \trans(\state[1][][\player,] \mid \state[0][][\player,], \action[][][0][\player,])\left[ \reward[\player]\left(\state[1][][\player,], \action[][][1][\player,]; \param \right) + \sum_{\iter = 2}^\infty \reward[\player]\left(\state[\iter][][\player,], \action[][][\iter][\player,] \right) \prod_{k = 2}^{\iter} \discount^{k+1} \trans(\state[k][][\player,] \mid \state[k-1][][\player,], \action[][][k-1][\player,]) \right] \right]
\end{align}
\begin{align}
    \obj[\param](\param, \otherstrat; \hist, \hist[][\prime]) &\doteq \sum_{\player \in \players} \left[\sum_{\iter}\grad[\param]\reward[\player](\state[\iter][][\player,], \action[][][\iter][\player,]; \param)  -   \sum_{\iter} \grad[\param] \reward[\player](\state[\iter][][\prime], \action[][][\iter][\prime]; \param) \right]
\end{align}

Under \Cref{assum:smooth_convex_invex}, these estimators are unbiased estimates of the gradients $\grad[\param] \obj$ and $\grad[\strat] \obj$, respectively.
Assuming these estimators have bounded variance, we can now solve the min-max optimization problem $\min_{\param \in \params} \max_{\strat \in \stratspace} \obj(\param, \strat)$ for an $\varepsilon$-inverse NE in $O(\nicefrac{1}{\varepsilon})$ iterations via stochastic gradient descent (\Cref{alg:online-sgda}).%
\footnote{With additional care, the assumption made on the rewards and probability transition functions in Part 2 of \Cref{assum:smooth_convex_invex} can be weakened to continuous differentiability and local Lipschitz-continuity, respectively (see
Lemma 3.2 of \citet{suh2022differentiable}) to obtain unbiased estimates;
for clarity we make the stronger assumption.} 
It thus remains to show that the variance of the gradient estimators are bounded: i.e., there exists $\variance \in [0, \infty)$ s.t.\@ for all $(\param, \strat) \in \params \times \stratspace$, $\left\| \Ex_{\hist, \hist[][\prime]}[(\obj[\param], \obj[\strat]) (\param, \strat; \hist, \hist[][\prime])] - \grad \obj(\param, \strat) \right\| \leq \variance$. 
Since rewards, transitions, and policies, are twice continuously-differentiable, both the gradient estimates $(\obj[\param], \obj[\strat])$ and $\grad \obj$ also are, and we have:
$\left\|\Ex_{\hist, \hist[][\prime]}[(\obj[\param], \obj[\strat])(\param, \strat; \hist, \hist[][\prime])] - \grad \obj(\param, \strat) \right\|\leq \max_{\param, \strat, \hist, \hist[][']} \left\|  (\obj[\param], \obj[\strat])(\param, \strat; \hist, \hist[][\prime])\right\| = \left\|(\obj[\param], \obj[\strat])\right\|_\infty$, where the max is well defined, since the objective is continuous and the maximization domains $\states, \actionspace, \stratspace, \params$ are non-empty and compact. 
This means that, under \Cref{assum:smooth_convex_invex}, the variance of gradient estimator $(\obj[\param], \obj[\strat])$ is bounded by $\variance^2 \doteq \left\|(\obj[\param], \obj[\strat])\right\|_\infty^2$.

\end{document}